\documentclass[aps,superscriptaddress,nofootinbib,eqsecnum,prd,notitlepage,
showkeys]{revtex4-1} 

\pdfoutput=1

\usepackage{amsfonts}
\usepackage{amsmath}
\usepackage{amssymb}
\usepackage{graphicx,color}
\usepackage{float}
\usepackage{hyperref}
\usepackage{subfigure}
\usepackage{dcolumn}
\usepackage{soul}
\usepackage{ulem}
\usepackage{verbatim}

\begin{document}

\title{Behaviour of $\alpha$-attractors in Warm Inflation}

\author{Dibya Chakraborty}
\email{dibyac@physics.iitm.ac.in}
\affiliation{Department of Physics, Ashoka University,
   Rajiv Gandhi Education City, Rai, Sonipat: 131029, Haryana, India}
\affiliation{Centre for Strings, Gravitation and Cosmology, Department of Physics, Indian Institute of Technology Madras, Chennai 600036, India}

\author{Suratna Das}
\email{suratna.das@ashoka.edu.in}
\affiliation{Department of Physics, Ashoka University,
   Rajiv Gandhi Education City, Rai, Sonipat: 131029, Haryana, India}

\begin{abstract}

The $\alpha$-attractor models of inflation have remained one of the preferred inflationary models for nearly a decade now. The unique attractor nature of these models in the $n_s-r$ plane have put these models in the sweet-spot of the $n_s-r$ measurement of Planck observations. In this article, we analyse the behaviour of such attractor models in a Warm Inflation setup to investigate whether the attractor nature of these models can be retained even when the dynamics deviates from the standard Cold inflationary dynamics. We have chosen to analyse these models in a strongly dissipative Warm inflation setup, namely the Minimal Warm Inflation, as in such a setup the inflationary dynamics significantly departs from the standard Cold inflation dynamics. We observe that the departure from the standard Cold inflation dynamics destroys the unique attractor nature of such models. The analysis clearly indicates that the attractor nature of these $\alpha$-attractor models is quite unique to the standard Cold inflationary dynamics. However, on a positive note, the analysis indicates that the $\alpha$-attractor models may be made in tune with  the recent ACT results in Warm inflation for certain parameter ranges. 

\end{abstract}

\maketitle

\section{Introduction}

Planck's latest measurement of the primordial scalar spectral index ($n_s$) as $n_s=0.9651\pm0.0044$ \cite{Planck:2018vyg} and the upper bound set on tensor-to-scalar ratio $r$ by the joint analysis of Planck and BICEP/Keck as $r<0.036$ \cite{BICEP:2021xfz} have put stringent bounds on the inflationary observables disfavouring quite a few popular inflationary models including the simple chaotic inflationary potentials. On the other hand, the $\alpha$-attractor models proposed in 2013 \cite{Kallosh:2013yoa} and its variants \cite{Kallosh:2013hoa, Kallosh:2022feu} have remained one of the preferred models of inflation in the past decade as its attractor nature in the $n_s-r$ plane put these models in the sweet-spot of the Planck's $n_s-r$ plot. However, the recent Atacama Cosmology Telescope (ACT) data along with Planck data and the BAO data of the DESI observation have shifted the value of the scalar spectral index towards unity ($n_s=0.9743\pm0.0034$) \cite{ACT:2025fju, ACT:2025tim}. This ACT joint analysis has now shifted the lime light from the $\alpha$-attractor models as being the favoured ones \cite{Kallosh:2025ijd}. Yet, the attractor nature of these $\alpha$-attractor models in the $n_s-r$ plane is quite unique and worth exploring in other variant inflationary scenarios. 

One of the variant inflationary scenarios of the standard inflationary paradigm is Warm Inflation (WI) \cite{Berera:1995ie}. In the standard inflationary picture, the inflaton field slowly rolls down its flat potential to yield inflation while the coupling of the inflaton field with other degrees of freedom is considered to be suppressed, and the inflaton field remains the sole dominant field during inflation. This leads to a universe devoid of any energy densities post inflation which then calls for a period of reheating through which the universe enters a radiation dominated epoch \cite{Riotto:2002yw, Baumann:2009ds}. We will refer to this standard inflationary paradigm as Cold Inflation (CI) henceforth. In contrast, the WI paradigm does consider the couplings to other degrees of freedom to be active during inflation, and the inflaton field dissipates its energies through these couplings to a subdominant, yet non-negligible, radiation energy density \cite{Kamali:2023lzq, Berera:2023liv}. The advantage of such a dynamics is that when WI ends, the inflationary phase smoothly transits to a radiation dominated epoch without any need of a separate reheating phase, physics of which is still largely unknown. WI has certain other advantages over CI. Firstly, WI produces less tensor-to-scalar ratios with respect to CI, which helps certain inflationary potentials, like the quartic self-interaction potential, to be in tune with the Planck data \cite{Bartrum:2013fia} that are otherwise ruled out for overproducing the tensor modes with respect to the scalar ones. It has recently been shown that chaotic quadratic potential can be made compatible with the recent ACT data in a WI setup \cite{Berera:2025vsu}. Secondly, WI can accommodate steeper potentials which are not valid inflationary potentials in CI setup \cite{Das:2020xmh}. Moreover, WI can comply with the distance and the de Sitter Swampland conjectures much more easily than CI due to its dynamics \cite{Das:2018hqy, Das:2018rpg, Motaharfar:2018zyb, Das:2019hto, Das:2019acf} and hence are better suited as an inflationary paradigm for ultraviolet complete theories, like String Theory, than CI. Recently, the fibre inflationary models constructed in type-IIB String Theory have been shown to be in tune with the Planck as well as the ACT results in a WI setup \cite{Chakraborty:2025yms}.  Above all, WI models yield distinct non-Gaussian features which can potentially distinguish such inflationary scenarios from the CI ones \cite{Bastero-Gil:2014raa, Mirbabayi:2022cbt}.

WI differs quite significantly from CI at the perturbation level as well. Due to the presence of the radiation bath during WI, not only the primordial quantum fluctuations are generated during WI but it also gives rise to classical thermal fluctuations which contribute dominantly to the primordial scalar power spectrum. Due to this fact, the scalar power spectrum in WI takes a more intricate form than the standard CI \cite{Hall:2003zp, Graham:2009bf, Bastero-Gil:2011rva, Ramos:2013nsa}. This also results in significant departure in the form of $n_s$ from CI. The scalar spectral index in WI and its running as well as running of running have been analytically analyzed in \cite{Das:2022ubr}. On the other hand, as the tensor modes are not affected by the thermal bath, the tensor power spectrum retains the same form in WI as in CI. However, as the scalar spectrum is modified in WI, it also affects the tensor-to-scalar ratio. This indicates that both $n_s$ and $r$ take different form in WI than in CI. This raises the question: how would the $\alpha$-attractor inflationary models behave in the $n_s-r$ plane in a WI setup? The main objective of this article is to investigate such a question that will reveal  whether such an attractor behaviour is only a feature of CI or not. 

In WI, as the inflaton field dissipates its energy to a sub-dominant radiation bath, an extra friction term arises in the dynamics of the inflaton field due to such dissipative effects apart from the conventional Hubble friction term that exists in CI. Depending on which one of these two frictional terms dominates inflaton's dynamics, WI can be categorised into two regimes: weak dissipative regime, when the friction due to dissipation much weaker than the background Hubble friction, and strong dissipative regime, when the dissipative frictional term, being more effective than the Hubble friction, dominates inflaton's evolution \cite{Kamali:2023lzq, Berera:2023liv}. It is then evident that the dynamics of WI is maximally in contrast with that of CI when WI takes place in a strong dissipative regime, and one can expect to observe the effects of WI dynamics in the $n_s-r$ plane of $\alpha$-attractors most significantly in a strong dissipative WI model. Therefore, in this article, we chose the Minimal Warm Inflation (MWI) \cite{Berghaus:2019whh} model as our working model of WI where WI takes place in strong dissipative regime. 

We have organised the rest of the article as follows. In Sec.~\ref{attractor-CI} we will discuss the attractor behaviour of three different $\alpha$-attractor models, namely the T-model, the E-model and the polynomial model. We will discuss in detail how the attractor behaviour in the $n_s-r$ plane of these three models can be analytically analysed, as we will later extend such an analysis in a WI setup to look for the behaviour of these three attractor models in the $n_s-r$ plane in WI. In Sec.~\ref{WI-summary}, we will briefly discuss the WI dynamics, its scalar and tensor power spectra as well as the forms of $n-s$ and $r$ keeping our focus on MWI, which will be our working model of WI. It is of importance to note that, unlike in CI, graceful exit from WI is an intricate process which depends on the form of the inflaton potential as well as the form of the dissipative term of the WI model \cite{Das:2020lut}. Demanding natural graceful exit from WI often puts constraints on the model parameters. However, in Sec.~\ref{graceful-exit} we will show that MWI naturally exits from inflationary phase when combined with the three type of $\alpha$-attractor potentials without putting any constraints on the model parameters of these three types of $\alpha$-attractor models. Our main analysis of this article will be furnished in Sec.~\ref{attractor-WI}. In  this section, we will first extend the analysis of the attractor behaviour of these $\alpha$-attractor models in CI presented in Sec.~\ref{attractor-CI} to WI, and later we will explore the behaviour of the three different attractor models in the $n_s-r$ plane one by one to show that how WI modifies the attractor behaviour of such models in a strong dissipative WI model. At the end, in Sec.~\ref{conclusion}, we will discuss the main results of our analysis and conclude by featuring some further outstanding issues that can be explored in the future.

\section{$\alpha$-attractor models: a brief overview}
\label{attractor-CI}

It has been observed that a large class of inflationary models, such as the Starobinsky model \cite{Starobinsky:1980te}, the quartic inflationary models with non-minimal coupling to gravity \cite{Bezrukov:2007ep,Okada:2010jf,Linde:2011nh}, various conformal, superconformal, and supergravity generalizations of these models \cite{Ellis:2013xoa,Buchmuller:2013zfa,Kallosh:2013hoa,Kallosh:2013daa,Kallosh:2013pby,Kallosh:2013lkr,Ellis:2013nxa,Ferrara:2013rsa,Kallosh:2013maa,Kallosh:2013tua}, lead to identical observational predictions for the scalar spectral index $n_s$ and the tensor-to-scalar ratio $r$ in the limit of large number of $e$-folds $N$ \cite{Roest:2013fha}:
\begin{eqnarray}
\label{general_ns_r}
n_s=1-\frac{2}{N}, \quad\quad r=\frac{12}{N^2}.
\end{eqnarray}
These models predict $0.96\lesssim n_s\lesssim 0.97$ and $0.003\lesssim r\lesssim 0.005$ for $50\leq N\leq 60$ which are in excellent agreement with the present observations 
\cite{Planck:2018jri}. Because of the universal nature of their observational predictions, these models are called {\it cosmological attractors}. Later in \cite{Kallosh:2013yoa}, the superconformal models of this class have been generalised by introducing a parameter $\alpha$ which is inversely proportional to the curvature of the inflaton K\"{a}hler manifold. It was shown that for sufficiently large curvature, i.e. small $\alpha$, the above mentioned predictions are modified as 
\begin{eqnarray}
\label{general_ns_r_alpha}
n_s=1-\frac{2}{N}, \quad\quad r=\alpha\frac{12}{N^2},
\label{predictions-alpha}
\end{eqnarray}
which are also favoured by the present data. These models are dubbed $\alpha$-attractors in the literature. However, in the small curvature limit, i.e. for large $\alpha$, such models behave like generic chaotic inflationary models. 

The reason behind this attractive nature in the $n_s-r$ plane of these models is that the potential of the inflaton field in the Einstein frame reaches a plateau in the large field limit in all these models. Such as, in the cosmological attractor models the potential assumes a general form in the Einstein frame as 
\begin{eqnarray}
V(\varphi)=V_0(1-e^{-\sqrt{\frac32}\varphi}), 
\end{eqnarray}
whereas in a generic $\alpha$-attractor model, the potential in the Einstein frame takes the form 
\begin{eqnarray}
V(\varphi)=V(\sqrt{6\alpha}\tanh(\varphi/\sqrt{6\alpha})).
\end{eqnarray}
Both of these potentials approach a plateau in the limit of large fields ($\varphi\rightarrow\infty$). Depending on the form of the potential and the rate at which they approach the plateau, $\alpha$-attractor models are often categorised in the following three types of model:

\subsection {T-model} 

This class of potentials was first proposed and analyzed in the context of {\it cosmological attractors} retaining conformal symmetry by Kallosh and Linde in \cite{Kallosh:2013hoa}. 
It was shown in \cite{Kallosh:2013hoa} that a two-field conformally invariant model,
\begin{eqnarray}
\mathcal{L}=\sqrt{-g}\left[\frac{1}{12}R(\chi^2-\phi^2)+\frac12\partial_\mu\chi\partial^\mu\chi-\frac12\partial_\mu\phi\partial^\mu\phi-\frac{1}{36}F(\phi/\chi)(\phi^2-\chi^2)^2\right],
\label{L-conformal}
\end{eqnarray}
where $F$ is an arbitrary function of ratio $\phi/\chi$ ($F=$ constant represents a model with $SO(1,1)$ de Sitter symmetry, whereas any arbitrary $F$ indicates breaking of the de Sitter symmetry), can be written after a conformal gauge fixing $\chi^2-\phi^2=6$, and redefinition of the fields $\chi=\sqrt6\cosh\frac{\varphi}{\sqrt6}$ and $\phi=\sqrt6\sinh\frac{\varphi}{\sqrt6}$ as
\begin{eqnarray}
{\mathcal L}=\sqrt{-g}\left[\frac12 R-\partial_\mu\varphi\partial^\mu\varphi-F\left(\tanh\frac{\varphi}{\sqrt6}\right)\right].
\label{generic-Lag}
\end{eqnarray}
In the large $\varphi$ limit ($\tanh\varphi\rightarrow\pm1$ and $F\rightarrow$ constant), the above Lagrangian asymptotically evolves towards a critical point where the de Sitter symmetry is restored. The simplest choice of the function $F$ as $F(\phi/\chi)=\lambda_n(\phi/\chi)^{2n}$, where $n$ is an integer, yields the potential for the redefined field $\varphi$ as 
\begin{eqnarray}
V(\varphi)=\lambda_n\tanh^{2n}(\varphi/\sqrt6).
\label{T-pot}
\end{eqnarray}
This model was called the T-model in  \cite{Kallosh:2013hoa} as it originates from different powers of $\tanh(\varphi/\sqrt{6})$.

Such a two-field model can be realized in a Superconformal theory with three chiral supermultiplets: $X^0$ (called the `conformon'), $X^1=\Phi$ (playing the role of the inflaton) and $X^2=S$ (a sGoldstino). The Lagrangian of such a theory can be written as 
\begin{eqnarray}
{\mathcal L}=\sqrt{-g}\left[-\frac16{\mathcal N}(X,\bar{X})R-G_{I\bar{J}}{\mathcal D}^\mu X^I{\mathcal D}_\mu \bar{X}^{\bar{J}}-G^{I\bar{J}}{\mathcal W}_I\bar{\mathcal W}_{\bar J}\right], \quad\quad\quad (I,\,\bar{I}={0,1,2})
\end{eqnarray}
where the K\"{a}hler potential ${\mathcal N}(X,\bar{X})$ takes the form 
\begin{eqnarray}
{\mathcal N}(X,\bar{X})=-|X^0|^2+|X^1|^2+|S|^2,
\end{eqnarray}
whereas the superpotential $\mathcal{W}(X)$ can be written as
\begin{eqnarray}
{\mathcal W}=Sf(X^1/X^0)\left[(X^0)^2-(X^1)^2\right].
\end{eqnarray}
where $f$ is an arbitrary function of the ratio $X^1/X^0$. It was shown in \cite{Kallosh:2013hoa}  that upon stabilizing the extra moduli fields and stabilization of the sGoldstino at $S=0$, the inflationary Lagrangian can be written as 
\begin{eqnarray}
{\mathcal L}=\sqrt{-g}\left[\frac12 R-(\partial\varphi)^2-f\left(\tanh\frac{\varphi}{\sqrt6}\right)\right],
\end{eqnarray}
which resembles the form of the Lagrangian previously obtained in Eq.~(\ref{generic-Lag}). Here, $\varphi$ arises after the field redefinitions $X^0=\sqrt{6}\cosh(\varphi/\sqrt{6})$ and  $X^1=\sqrt{6}\sinh(\varphi/\sqrt{6})$.

\begin{center}
\begin{figure}[!htb]
\subfigure[ Varying $n$ with $\alpha=\frac73$ ]{\includegraphics[width=8cm]{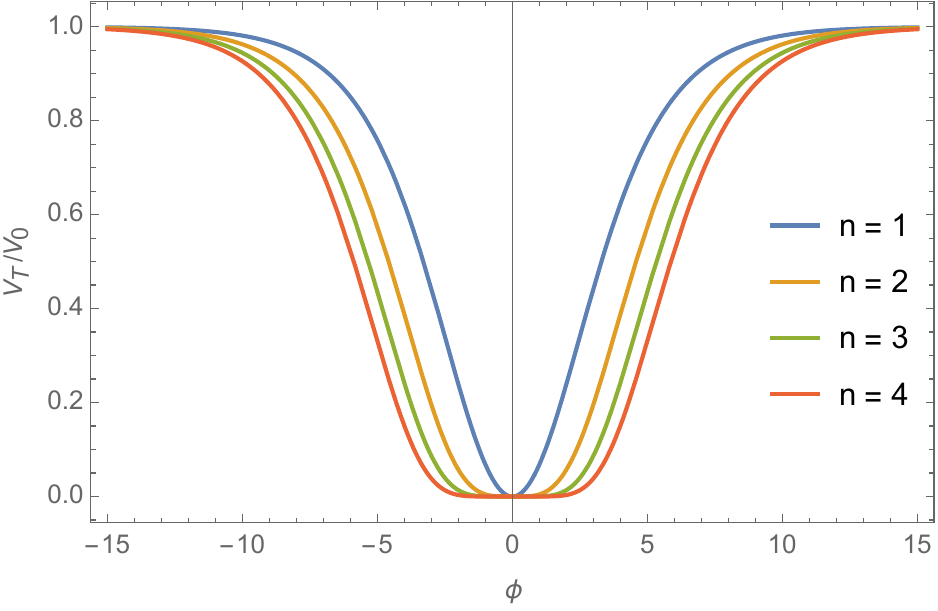}}
\subfigure[ Varying $\alpha$ with $n=3$ ]{\includegraphics[width=8cm]{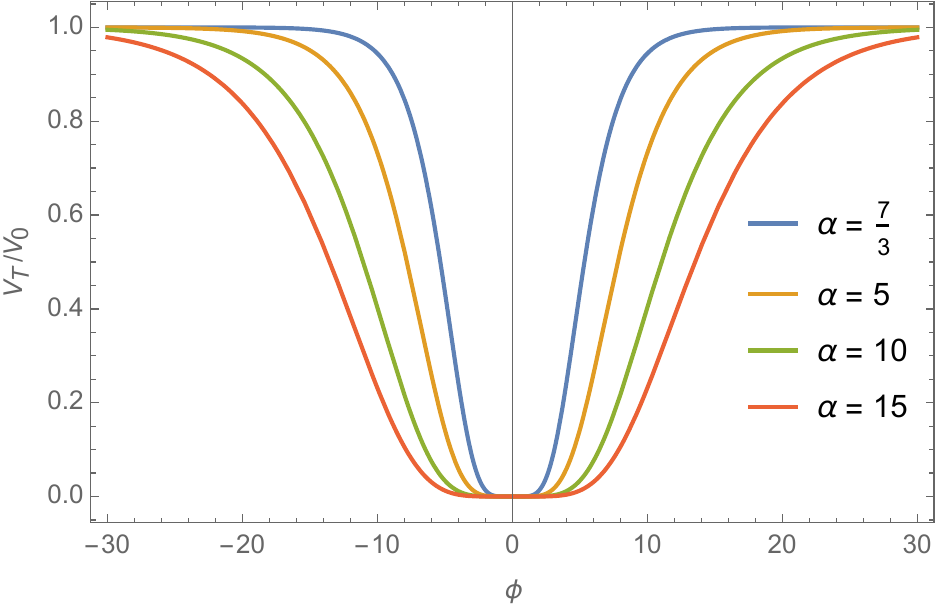}}
\caption{The T-model potential of $\alpha$-attractor models:  $V_T=V_0\tanh^{2n}(\varphi/\sqrt{6\alpha})$}
\label{T-model-pot}
\end{figure}
\end{center}

Such a superconformal theory was later generalized in \cite{Kallosh:2013yoa} by Kallosh and Linde by introducing a parameter $\alpha$ in the K\"{a}hler potential given as 
\begin{eqnarray}
{\mathcal N}(X,\bar{X})=-|X^0|^2\left[1-\frac{|X^1|^2+|S|^2}{|X^0|^2}\right]^\alpha,
\end{eqnarray}
and the superpotential takes the form 
\begin{eqnarray}
{\mathcal W}=S(X^0)^2f(X^1/X^0)\left[1-\frac{(X^1)^2}{(X^0)^2}\right]^{(3\alpha-1)/2}.
\end{eqnarray}
The parameter $\alpha$ turns our to be inverse of the constant curvature of the K\"{a}hler manifold $R_K=-g_{\Phi\bar\Phi}^{-1}\partial_\Phi\partial_{\bar\Phi}\log g_{\Phi\bar\Phi}$:
\begin{eqnarray}
\alpha=-\frac{2}{3R_K}
\end{eqnarray}
This generalized theory, upon moduli stabilization, yields an effective Lagrangian \cite{Kallosh:2013yoa}:
\begin{eqnarray}
{\mathcal L}=\sqrt{-g}\left[\frac12 R-\frac{\alpha}{(1-\Phi^2/3)^2}(\partial\Phi)^2-f^2(\Phi/\sqrt{3})\right].
\label{alpha-lag1}
\end{eqnarray}
A further field redefinition such as $\Phi/\sqrt3=\tanh(\varphi/\sqrt{6\alpha})$ will yield a Lagrangian of a scalar field with canonical kinetic term:
\begin{eqnarray}
{\mathcal L}=\sqrt{-g}\left[\frac12 R-(\partial\varphi)^2-f\left(\tanh\frac{\varphi}{\sqrt{6\alpha}}\right)\right].
\label{alpha-lag2}
\end{eqnarray}
This simplest class of superconformal models represents a general family of $\alpha$-attractors. 
As per the previous discussions, the simplest choice of the function $f$ leads to an inflaton potential of the form 
\begin{eqnarray}
V_T=V_0\tanh^{2n}(\varphi/\sqrt{6\alpha}),
\label{T-alpha}
\end{eqnarray}
which resembles the potential of the T-model given in Eq.~(\ref{T-pot}). The form of the potential is depicted in Fig.~(\ref{T-model-pot})

From the above discussions one can see that the analysis of the $\alpha$-attractor inflationary models is essentially equivalent to the analysis of a canonical single-field inflationary model with different choices of the functional forms of $f\left(\tanh\frac{\varphi}{\sqrt{6\alpha}}\right)$ which yield different forms of inflaton potentials. Hence, the standard results of a canonical single-field inflationary model hold for the $\alpha$-attractor models as well. Therefore, with the knowledge of the $\alpha$-attractor potential, one can derive the scalar spectral index $n_s$ as well as the tensor-to-scalar ratio $r$ using the standard results of single-field inflationary models:
\begin{eqnarray}
n_s=1-6\epsilon_V+2\eta_V,\quad\quad\quad r=16\epsilon_V,
\end{eqnarray}
where $\epsilon_V$ and $\eta_V$ are the standard potential slow-roll parameters:
\begin{eqnarray}
\epsilon_V=\frac{M_{\rm Pl}^2}{2}\left(\frac{V_{,\phi}}{V}\right)^2, \quad\quad\quad \eta_V=M_{\rm Pl}^2\left(\frac{V_{,\phi\phi}}{V}\right),
\end{eqnarray}
$\phi$ being the canonical inflaton field, $M_{\rm Pl}$ is the reduced Planck mass. 
The scalar spectral index and the tensor-to-scalar ratio are measured from the observational data when the pivot scale of wavemunber $0.05$ Mpc$^{-1}$ exits the horizon during inflation. The number of $e$-folds at the horizon exit of the pivot scale $N_*$ can also be calculated using the inflaton potential as 
\begin{eqnarray}
N_*=-\frac{1}{M_{\rm Pl}^2}\int_{\phi_*}^{\phi_e}d\phi\frac{V}{V_{,\phi}},
\label{Nstar-CI}
\end{eqnarray}
where $\phi_*$ and $\phi_e$ are the inflaton field values at the horizon-exit of the pivot scale and at the end of inflation, respectively. 

\begin{center}
\begin{figure}[!htb]
\includegraphics[width=14cm]{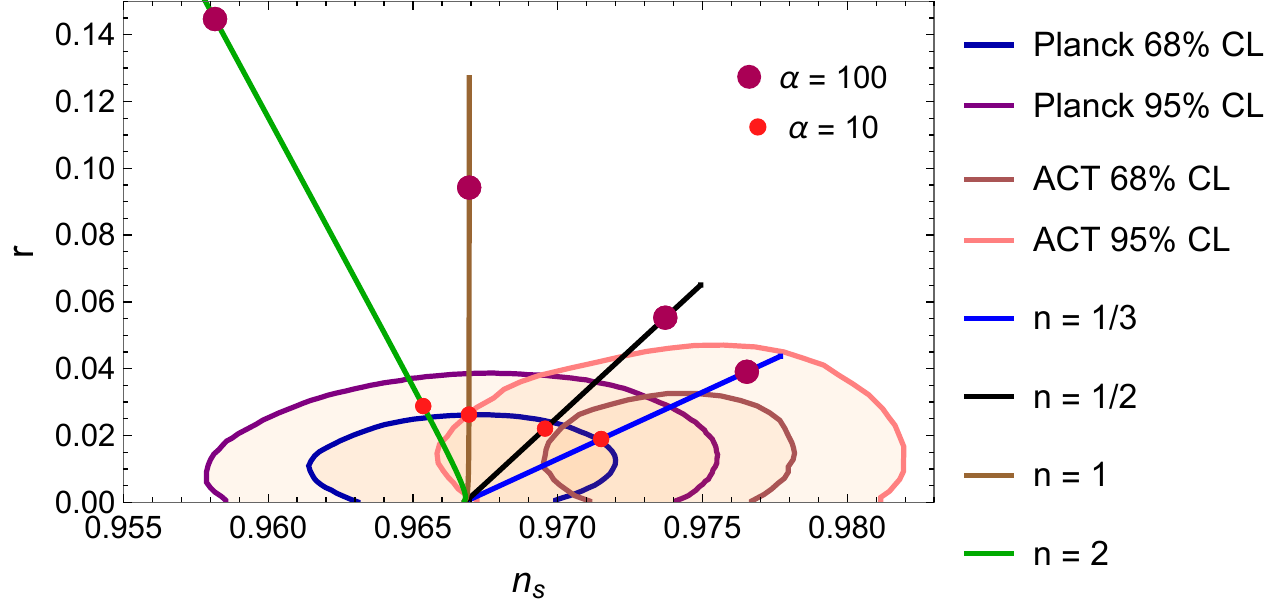}
\caption{The $n_s$ vs $r$ plot for T-model $\alpha$-attractor for $N_*=60$}
\label{T-model-nsr}
\end{figure}
\end{center}

Therefore, using the T-model $\alpha$-attractor potential given in Eq.~(\ref{T-alpha}), the slow-roll parameters take the form (setting $M_{\rm Pl}=1$):
\begin{eqnarray}
\epsilon_V=\frac{4n^2}{3\alpha\sinh^2(\sqrt{2/3\alpha}\,\phi)}, \quad\quad\quad \eta_V=\frac{8n^2-4n\cosh(\sqrt{2/3\alpha}\,\phi)}{3\alpha\sinh^2(\sqrt{2/3\alpha}\,\phi)}.
\end{eqnarray}
The field value at the end of inflation. $\phi_e$, can be calculated by setting $\epsilon_V=1$, yielding 
\begin{eqnarray}
\cosh(\sqrt{2/3\alpha}\,\phi_e)=\sqrt{\frac{4n^2+3\alpha}{3\alpha}},
\end{eqnarray}
which then provides the field value $\phi_*$ at $N_*$ as  
\begin{eqnarray}
\cosh(\sqrt{2/3\alpha}\,\phi_*)=\sqrt{\frac{4n^2+3\alpha}{3\alpha}}+\frac{4nN_*}{3\alpha}.
\end{eqnarray}
It is, therefore, straightforward to calculate $n_s$ and $r$ at the horizon-crossing of the pivot scale as 
\begin{eqnarray}
n_s&=&\frac{1-\frac{2}{N_*}-\frac{3\alpha}{4N_*^2}+\frac{1}{2nN_*}\left(1-\frac{1}{N_*}\right)g(\alpha,n)}{1+\frac{1}{2nN_*}g(\alpha,n)+\frac{3\alpha}{4N_*^2}},\\
r&=&\frac{12\alpha}{N_*^2+\frac{N_*}{2n}g(\alpha,n)+\frac34\alpha},
\end{eqnarray}
where $g(\alpha,n)\equiv\sqrt{3\alpha(4n^2+3\alpha)}$. The same results were obtained in \cite{Kallosh:2013yoa}. One can see that in the $\alpha\rightarrow0$ limit, this model predicts the $\alpha$-attractor results as conveyed in Eq.~(\ref{predictions-alpha}). Fig.~\ref{T-model-nsr} shows the $n_s$ vs $r$ plot for T-model $\alpha$-attractor setting $N_*=60$ which clearly depicts the attractor behaviour of such a model  in the limit $\alpha\rightarrow0$. 

\subsection{E-model}

\begin{center}
\begin{figure}[!htb]
\subfigure[ Varying $n$ with $\alpha=\frac73$ ]{\includegraphics[width=8cm]{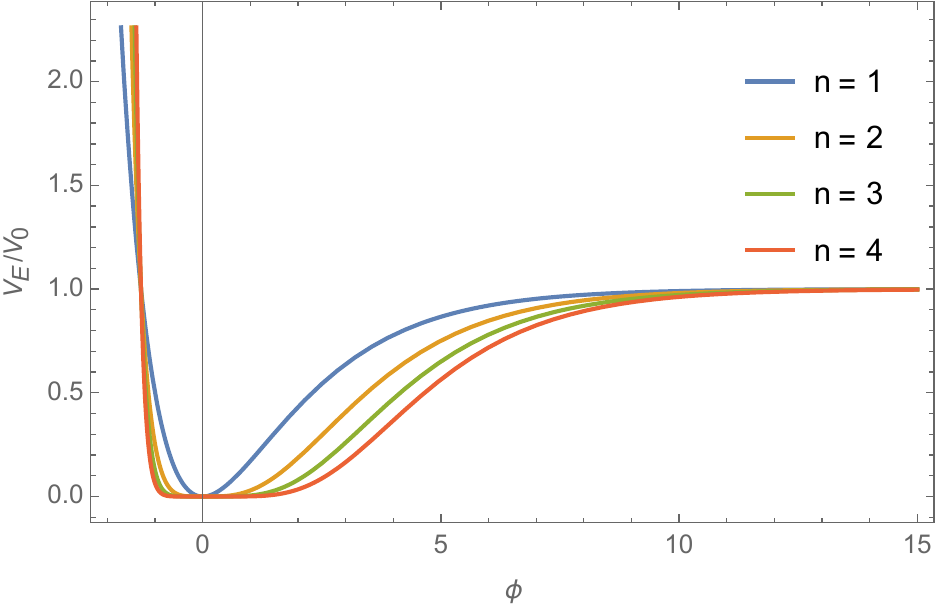}}
\subfigure[ Varying $\alpha$ with $n=3$ ]{\includegraphics[width=8cm]{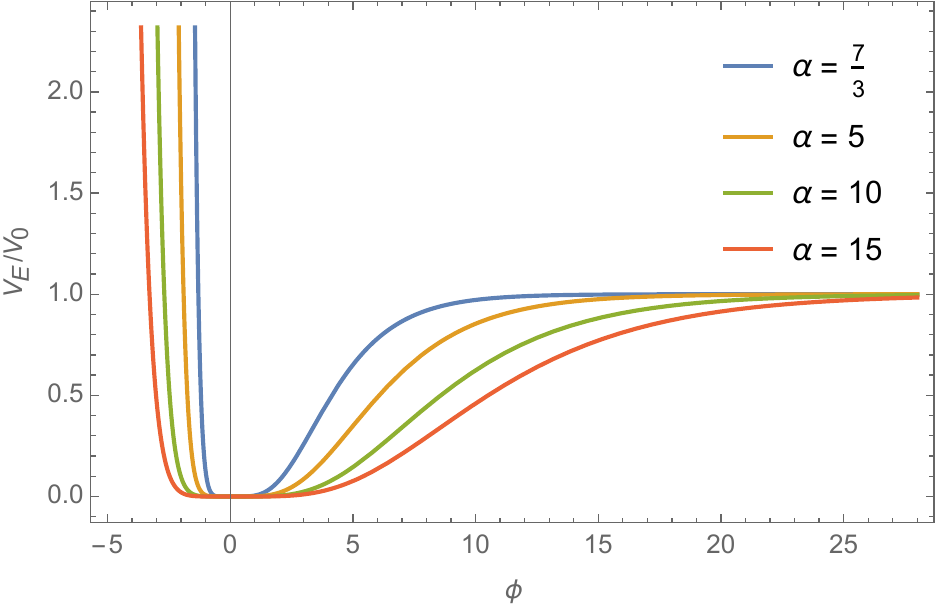}}
\caption{The E-model potential of $\alpha$-attractor models:  $V_E=V_0\left(1-e^{-\sqrt{2/3\alpha}\varphi}\right)^{2n}$}
\label{E-model-pot}
\end{figure}
\end{center}

In addition to the T-model, which arises with the simplest choice of the function $F(\phi/\chi)$ in Eq.~(\ref{L-conformal}), it was shown by Kallosh and Linde in \cite{Kallosh:2013hoa} that Starobinsky inflationary potentials can be recovered in Einstein frame by choosing $F(\phi/\chi)\sim\phi^2/(\phi+\chi)^2$, which yields 
\begin{eqnarray}
V(\varphi)\sim\left[\frac{\tanh(\varphi/\sqrt6)}{1+\tanh(\varphi/\sqrt6)}\right]^2\sim\left(1-e^{-\sqrt{2/3}\varphi}\right)^2.
\end{eqnarray}
A generalization of the form $F(\phi/\chi)\sim\phi^{2n}/(\phi+\chi)^{2n}$ will lead to the potential 
\begin{eqnarray}
V(\varphi)\sim\left(1-e^{-\sqrt{2/3}\varphi}\right)^{2n},
\end{eqnarray}
and further generalizing it to $\alpha$-attractors as before, one obtains 
\begin{eqnarray}
V_E(\varphi)=V_0\left(1-e^{-\sqrt{2/3\alpha}\varphi}\right)^{2n}.
\label{E-alpha}
\end{eqnarray}
For obvious reasons, such attractor inflationary potentials are dubbed E-model. The form of the $E$-model inflaton potential is depicted in Fig.~\ref{E-model-pot}.

\begin{center}
\begin{figure}[!htb]
\includegraphics[width=14cm]{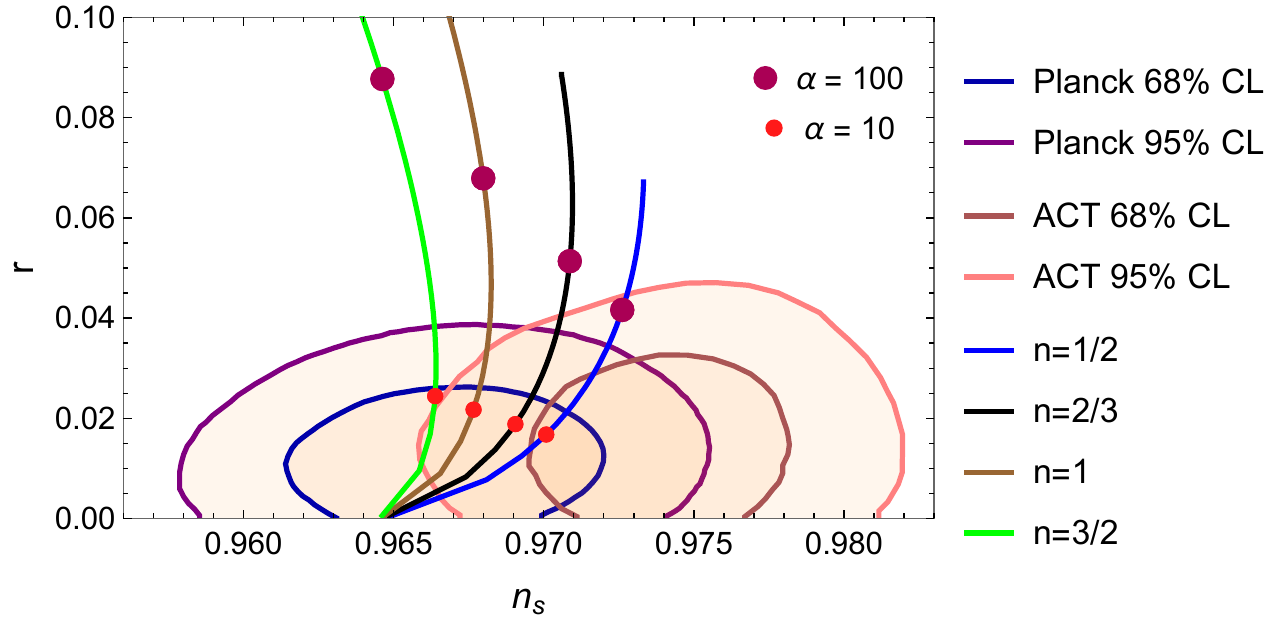}
\caption{The $n_s$ vs $r$ plot for E-model $\alpha$-attractor for $N_*=56$}
\label{E-model-nsr}
\end{figure}
\end{center}

Following the usual steps we find the potential slow-roll parameters as 
\begin{eqnarray}
\epsilon_V=\frac{4n^2}{3\alpha\left(e^{\sqrt{2/3\alpha}\phi}-1\right)^2},\quad\quad\quad \eta_V=\frac{8n^2-4ne^{\sqrt{2/3\alpha}\phi}}{3\alpha\left(e^{\sqrt{2/3\alpha}\phi}-1\right)^2},
\end{eqnarray}
and the inflaton field values $\phi_e$ and $\phi_*$ as \cite{Ballardini:2024ado}
\begin{eqnarray}
\sqrt{\frac{2}{3\alpha}}\phi_e&=&\ln\left[1+\frac{2n}{\sqrt{3\alpha}}\right],\nonumber\\
\sqrt{\frac{2}{3\alpha}}\phi_*&=&-\frac{4nN_*}{3\alpha}-\left(1+\frac{2n}{\sqrt{3\alpha}}\right)+\ln\left[1+\frac{2n}{\sqrt{3\alpha}}\right]-W_{-1}\left[-e^{-\frac{4nN_*}{3\alpha}-\left(1+\frac{2n}{\sqrt{3\alpha}}\right)+\ln\left[1+\frac{2n}{\sqrt{3\alpha}}\right]}\right],
\end{eqnarray}
where $W_{-1}$ is the product logarithm or the Lambert $W$ function. In this model, $n_s$ and $r$ take the form:
\begin{eqnarray}
n_s&=&1+\frac{8n}{3\alpha}\frac{W_{-1}\left[-e^{-\frac{4nN_*}{3\alpha}-\left(1+\frac{2n}{\sqrt{3\alpha}}\right)+\ln\left[1+\frac{2n}{\sqrt{3\alpha}}\right]}\right]-n}{\left(W_{-1}\left[-e^{-\frac{4nN_*}{3\alpha}-\left(1+\frac{2n}{\sqrt{3\alpha}}\right)+\ln\left[1+\frac{2n}{\sqrt{3\alpha}}\right]}\right]+1\right)^2},\\
r&=&\frac{64n^2}{3\alpha}\frac{1}{\left(W_{-1}\left[-e^{-\frac{4nN_*}{3\alpha}-\left(1+\frac{2n}{\sqrt{3\alpha}}\right)+\ln\left[1+\frac{2n}{\sqrt{3\alpha}}\right]}\right]+1\right)^2}.
\end{eqnarray}
Due to the presence of the product logarithms in the above expressions, they do not analytically tend to the standard attractor behaviour of $\alpha$-attractors in the $\alpha\rightarrow0$ limit as given in Eq.~(\ref{predictions-alpha}). However, the attractor behaviour is apparent in the $n_s-r$ plot as shown in Fig.~\ref{E-model-nsr}.


\subsection{Polynomial model}

\begin{center}
\begin{figure}[!htb]
{\includegraphics[width=12cm]{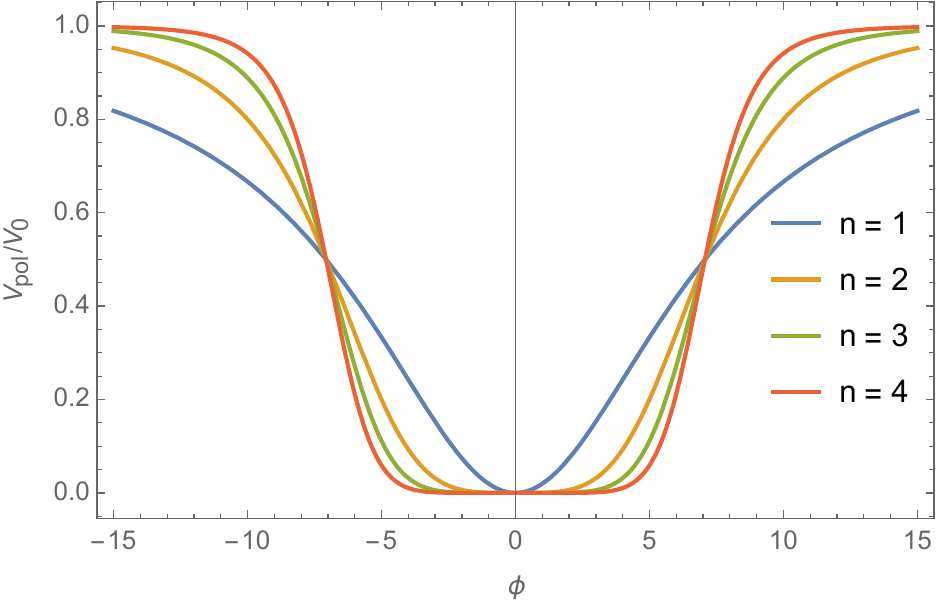}}
\caption{The E-model potential of $\alpha$-attractor models:  $V_P=V_0\frac{\phi^{2n}}{\phi^{2n}+\mu^{2n}}$ $\left(\alpha=\frac13,\, c=10,\, \mu\sim7.07\right)$}
\label{P-model-pot}
\end{figure}
\end{center}

\begin{center}
\begin{figure}[!htb]
\includegraphics[width=14cm]{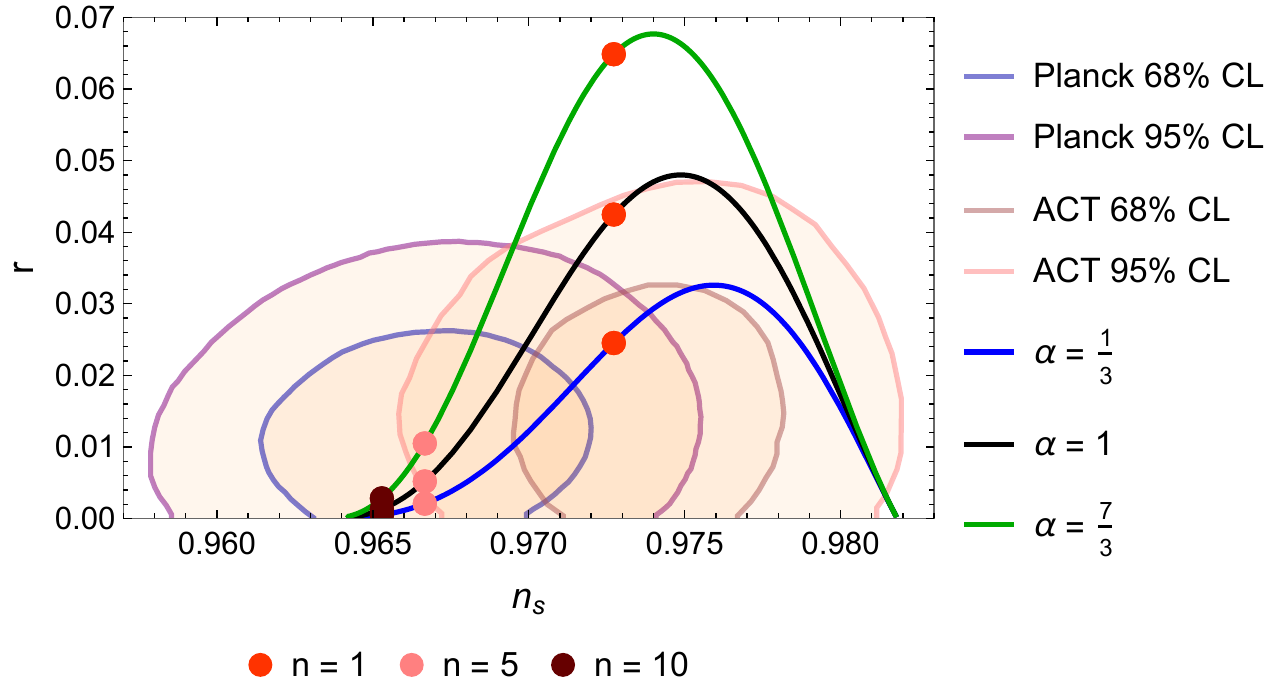}
\caption{The $n_s$ vs $r$ plot for polynomial model for $N_*=55$}
\label{pol-model-nsr}
\end{figure}
\end{center}

The polynomial $\alpha$-attractor models was introduced much later by Kallosh and Linde in \cite{Kallosh:2022feu}. In view of the previous analysis of the $\alpha$-attractor models (as given in Eq.~(\ref{alpha-lag1})) and Eq.~(\ref{alpha-lag2})), Kallosh and Linde studied the following Lagrangian in \cite{Kallosh:2022feu}
\begin{eqnarray}
\mathcal{L}=-\frac{3\alpha}{4}\frac{(\partial\rho)^2}{\rho^2}-V(\rho),
\end{eqnarray}
where the minimally coupled field $\rho$ is related to the canonical field $\phi$ as 
\begin{eqnarray}
\rho=\rho_0e^{-\sqrt{2/3\alpha}\phi}.
\end{eqnarray}
They chose a logarithmic potential for the field $\rho$ as
\begin{eqnarray}
V=V_0\frac{\ln^2\rho}{\ln^2\rho+c^2},
\end{eqnarray}
which yields a polynomial potential for the canonical field $\phi$:
\begin{eqnarray}
V=V_0\frac{\phi^2}{\phi^2+\mu^2},\quad\quad\quad \mu=\sqrt{\frac{3\alpha}{2}}c.
\end{eqnarray}
Upon generalization of this potential, one gets the polynomial $\alpha$-attractor potential \cite{Kallosh:2022feu} as
\begin{eqnarray}
V_P=V_0\frac{\phi^{2n}}{\phi^{2n}+\mu^{2n}}.
\end{eqnarray}
The form of the potential is shown in Fig.~\ref{P-model-pot}
As we can see from Fig.~\ref{P-model-pot}  that the polynomial $\alpha$-attractor potential reaches a plateau in the large field limit $\phi>\mu$, we will study this model in this region. The slow-roll parameter in this large field limit turn out to be 
\begin{eqnarray}
\epsilon_V&=&\frac{2n^2}{\mu^2}\left(\frac{\mu}{\phi}\right)^{4n+2}\left(1-2\left(\frac{\mu}{\phi}\right)^{2n}\right),\nonumber\\
\eta_V&=&-\frac{2n(1+n)}{\mu^2}\left(\frac{\mu}{\phi}\right)^{2n+2}\left(1-2\left(\frac{\mu}{\phi}\right)^{2n}\right)\left(1-\frac{1-2n}{1+2n}\left(\frac{\mu}{\phi}\right)^{2n}\right).
\label{slow-roll-param-poly}
\end{eqnarray}
It is to note that in the limit $\phi\gg\mu$, one has $|\eta_V|\gg\epsilon_V$. We also get the inflaton field values $\phi_e$ and $\phi_*$ as 
\begin{eqnarray}
\frac{\phi_e}{\mu}&=&2^{\frac1{2+4n}}\left(\frac{\mu^2}{n^2}\right)^{-\frac1{2+4n}},\nonumber\\
\frac{\phi_*}{\mu}&=&\left(\frac{2}{\mu}\right)^{\frac1{1+n}}\left(n(n+1)N_*+2^{\frac{1+n}{1+2n}}n^{\frac{2+2n}{1+2n}}\mu^{\frac{2n}{1+2n}}\right)^{\frac1{2+2n}},
\end{eqnarray}
which then yield the $n_s$ and $r$ as \cite{Dimopoulos:2016zhy, Adhikari:2020xcg, Bhattacharya:2022akq}
\begin{eqnarray}
n_s&=&1-\frac{2}{N_*}\frac{2n+1}{2n+2},\\
r&=&\frac{2^{\frac{n+3}{n+1}}(\mu^{2n}n)^{\frac{1}{1+n}}}{\left((n+1)N_*\right)^{\frac{2n+1}{n+1}}}.
\end{eqnarray}
Here we note that, unlike the previous two cases, in this case, $n_s$ is  independent of the parameter $\alpha$ (through the parameter $\mu$). Thus, the polynomial model doesn't have an attractor behaviour in the parameter $\alpha$. Rather, if we take $n\rightarrow\infty$ limit, we see that \cite{Bhattacharya:2022akq}
\begin{eqnarray}
n_s\sim 1-\frac{2}{N_*}, 
\end{eqnarray}
and the scenario shows an attractor behaviour in the $n_s-r$ plane (one also should note that $n\rightarrow 0$ limit would produce $n_s\sim1-1/N_*$ which, though can behave as an atractor, is not observationally viable). In view of this, it seems that addressing the polynomial models an $\alpha-$attractor is a misnomer as there's no attractor behaviour in the $\alpha$ parameter. We show the attractor behaviour of such polynomial models in Fig.~\ref{pol-model-nsr}.
\section{A summary of essentials of Warm Inflation}
\label{WI-summary}

In Warm Inflation, the inflaton field, $\phi$, continuously dissipates its energy to a subdominant radiation bath, $\rho_r$, which remains in near thermal equilibrium so that a temperature $T$ can be defined of that radiation fluid. In this way, a constant subdominant radiation energy density is maintained throughout the evolution of WI that helps the inflationary system gracefully exit in a radiation dominated epoch. Due to the presence of such a dissipative term, the evolution of the inflaton field and the subdominant radiation fluid are coupled and can be expressed as 
\begin{eqnarray}
&&\ddot\phi+3H\dot\phi+V_{,\phi}=-\Upsilon(\phi,T)\dot\phi,\label{eom-phi} \\
&&\dot\rho_r+4H\rho_r=\Upsilon(\phi,T)\dot\phi^2 \label{eom-rhor}. 
\end{eqnarray}
Here, $\Upsilon(\phi,T)$ is the dissipative term through which the inflaton transfers its energy to the radiation bath. In general, one can write a generic form of the dissipative term $\Upsilon(\phi,T)$ arising in various WI models as 
\begin{eqnarray}
\Upsilon(\phi,T)=C_\Upsilon M^{1-p-c}T^p\phi^c,
\label{dis-coeff}
\end{eqnarray}
where $C_\Upsilon$ is a constant whereas $M$ is some appropriate mass scale. It is to note that the dissipative term in the inflaton's equation of motion (as given in Eq.~(\ref{eom-phi})) gives rise to an extra friction term ($\Upsilon\dot\phi$) apart from the one present due to the background expansion ($3H\dot\phi$). Defining a dimensionless quantity $Q$ which is the ratio of these two friction terms present in the inflaton's equation of motion as
\begin{eqnarray}
Q=\frac{\Upsilon}{3H},
\end{eqnarray}
one can classify WI scenarios into two categories: 1) The regime when $Q\ll1$, i.e. when the friction due to the background expansion dominates the inflaton's evolution, is called weak dissipative regime, and 2) the regime when $Q\gg1$, i.e.  the dissipative friction term dominates the inflaton's evolution, is called strong dissipative regime. It is obvious that the dynamics of WI taking place in weak dissipation is closer to that of CI, whereas the dynamics of WI in strong dissipative regime is quite different from CI. 
Moreover, the radiation fluid in thermal equilibrium can be written as 
\begin{eqnarray}
\rho_r=\frac{\pi^2}{30}g_*T^4\equiv C_rT^4,
\end{eqnarray}
where $g_*$ accounts for the relativistic degrees of freedom consisting the thermal bath. 
WI also differs from CI at the perturbation level because in WI there are thermal fluctuations present besides the usual quantum fluctuations. Therefore, the scalar power spectrum in WI takes the form \cite{Hall:2003zp, Graham:2009bf, Bastero-Gil:2011rva, Ramos:2013nsa}
\begin{eqnarray}
{\mathcal P}_{\mathcal R}=\left(\frac{H^2}{2\pi\dot\phi}\right)^2\left(1+2n_{\rm BE}+\frac{2\sqrt3\pi Q}{\sqrt{3+4\pi Q}}\frac{T}{H}\right)G(Q),
\label{scalar-ps}
\end{eqnarray}
where $n_{\rm BE}$ describes the Bose-Einstein distribution of the inflaton fluctuations if the inflaton field thermalizes with the concurrent radiation bath, and $G(Q)$ is known as the growth factor that appears due the coupling of the inflaton fluctuations with the thermal fluctuations and its form depends on the inflaton potential as well as the form of the dissipative coefficient of the WI model. On the other hand, as the tensor modes don't couple to the thermal fluctuations in WI, the tensor spectrum takes the same form as in CI:
\begin{eqnarray}
{\mathcal P}_T=\frac{H^2}{2\pi^2M_{\rm Pl}^2}. 
\end{eqnarray} 
The tensor-to-scalar ratio $r$ in WI can be determined as 
\begin{eqnarray}
r={\mathcal P}_T/{\mathcal P}_{\mathcal R}.
\label{r-WI}
\end{eqnarray}
The reader may refer to the two recent reviews on WI for more details \cite{Kamali:2023lzq, Berera:2023liv}. 

To analyse the attractor behaviour of the $\alpha$-attractor models in WI, we will choose strong dissipative regime because in this regime the dynamics of WI maximally differs from that of CI. Thus the attractor behaviour will get affected the most in the strong dissipative WI models. Moreover, the strong dissipative regime has other advantages as follows:
\begin{enumerate}
\item The scalar power spectrum is not sensitive to the thermalization of the inflaton field and hence one can ignore the $2n_{\rm BE}$ factor in the scalar power spectrum in strong dissipative regime \cite{Ballesteros:2023dno, Kumar:2024hju}.
\item The growth factor $G(Q)$ very weakly depends on the form of the potential in strong dissipative regime \cite{Das:2020xmh}. 
\end{enumerate}
These two features of strong dissipative regime will further simplify our analysis as we will see below.

\subsection{A specific realization of strong dissipative WI model: The Minimal Warm Inflation}

Recently a concrete microscopic realization of WI has  been proposed in which WI generally takes place in strong dissipative regime ($Q\gg1$), a.k.a. the Minimal Warm Inflation \cite{Berghaus:2019whh}. In this specific model an axionic inflaton field couples to non-Abelian gauge fields, and the shift symmetry of the axion field protects the inflaton potential from any perturbative backreactions and from acquiring a large thermal mass. Therefore, in such a WI model thermal corrections do not potentially distort the flatness of the inflaton potential. The dissipation of the inflaton field occurs through classical sphelaron transitions which yields a dissipative coefficient of the form:
\begin{eqnarray}
\Upsilon(T)\propto \alpha_g^5\frac{T^3}{f^2},
\end{eqnarray}
where $f$ is the axion decay constant and $\alpha_g=g^2/(4\pi)$, $g$ being the Yang-Mills gauge coupling. Non-gaussian signatures of such a WI scenario where the inflaton potential is treated as a quartic one ($\lambda\phi^4$) has been extensively studied recently in \cite{Mirbabayi:2022cbt}. It was shown that the model in strong dissipative regime can yield $f_{\rm NL}\sim1$ in the equilateral configuration which  asymptotes to 4.5 when $Q\sim{\mathcal O}(10^3)$ whereas produces vanishingly small non-Gaussianities in the squeezed limit. On the other hand, as the inflaton field is coupled to a radiation bath in WI, the scalar perturbations will be sourced by a noise term dictated by the Fluctuation Dissipation Theorem. The presence of the source noise can give rise to considerable non-Gaussianities apart from the ones arising due to non-linearities in the perturbations. A new analytical template,``new-warm" shape, has been proposed in \cite{Mirbabayi:2022cbt} to detect the non-Gaussianties arising in such a WI model. A detection of such a new shape in bispectrum non-Gaussianity will be a smoking gun signature of WI dynamics. 

As mentioned above, we chose to analyse WI in strong dissipation to quantify the attractor behaviour of a Warm inflationary $\alpha$-attractor model. We thus choose the Minimal Warm Inflation (MWI) model \cite{Berghaus:2019whh}, described above, as our strong dissipative WI model where the dissipative term depends on the cubic power of the temperature and has no dependence on the inflaton field,
\begin{eqnarray}
\Upsilon_{\rm MWI}= C_{\Upsilon}T^3/M^2.
\end{eqnarray}
Such a choice of the WI model will help ease the calculations of $N_*$ in the $\alpha$-attractor models as we will see in the subsequent sections. However, in the Supergravity constructions of the $\alpha$-attractor models, the real part of a complex scalar plays the role of the inflaton field, whereas the phase part is an axion field which is stabilized \cite{Yamada:2018nsk}. This inflaton field is also protected by the shift symmetry (which may be slightly broken), and thus can avoid large loop-corrections to the inflaton potential maintaining the flatness of the potential \cite{Yamada:2018nsk}. Here, we will assume that the inflaton field of the Supergravity $\alpha$-attractor models is axion-like. We further assume that this axion-like inflaton field of the $\alpha$-attractor models has similar non-Abelian gauge couplings as in MWI, which will help the field dissipate to a thermal bath through sphelaron heating. Such gauge couplings will also protect the inflaton potential from acquiring a large thermal mass as it happens in MWI.

\begin{center}
\begin{figure}[!htb]
\includegraphics[width=15cm]{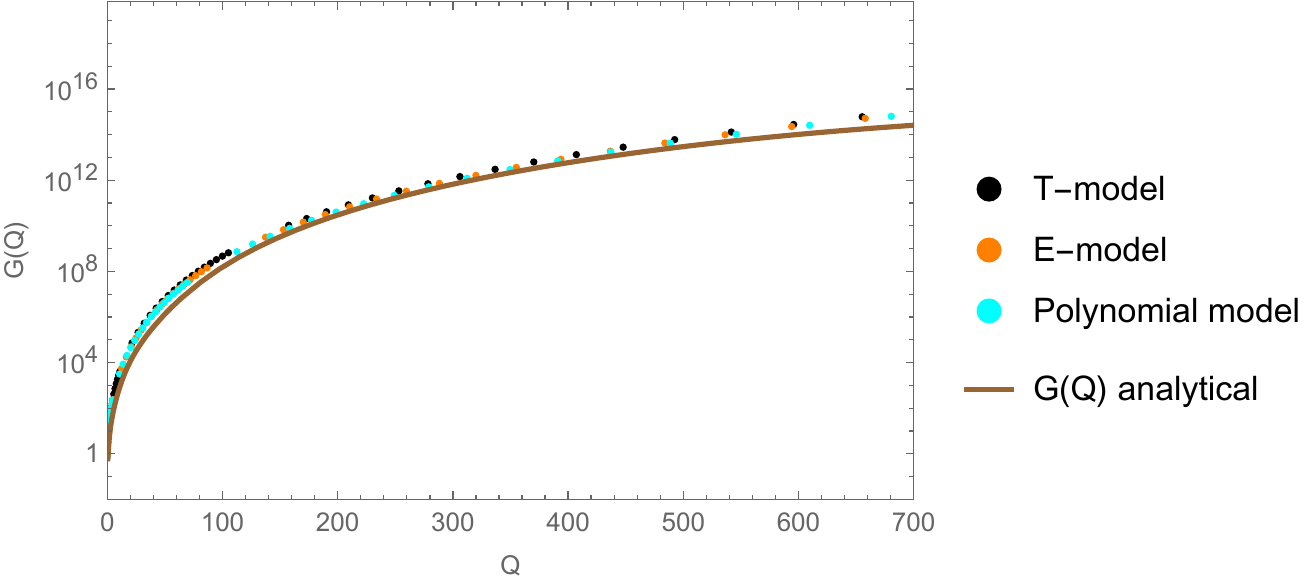}
\caption{The $n_s$ vs $r$ plot for T-model in WI for $N_*=51$}
\label{GQ-compare}
\end{figure}
\end{center}

To analyse the attractor behaviour of the Warm $\alpha$-attractor scenario, it is then left to determine the scalar power spectrum and the scalar spectral index of the MWI model in combination with the $\alpha$-attractor potentials. The original MWI paper did not formally determine the $G(Q)$ factor appearing in the scalar power spectrum, as in Eq.~(\ref{scalar-ps}). Rather, they determined and dealt with an approximate form of the scalar power spectrum. The growth factor, $G(Q)$, for such a model in strong dissipative regime was first determined in \cite{Das:2020xmh} for runaway potentials. However, as in strong dissipative regime $G(Q)$ depends very weakly on the form of the inflaton potential \cite{Rodrigues:2025neh}, we can safely use the same form of the growth factor, 
\begin{eqnarray}
G_{\rm MWI}(Q)=\frac{1+6.12Q^{2.73}}{(1+6.96Q^{0.78})^{0.72}}+\frac{0.01Q^{4.61}(1+4.82\times10^{-6}Q^{3.12})}{(1+6.83\times10^{-13}Q^{4.12})^2},
\end{eqnarray}
even for the $\alpha$-attractor potentials. Recently, couple of codes (WarmSPy \cite{Montefalcone:2023pvh} and WI2easy \cite{Rodrigues:2025neh}) were proposed in literature to numerically calculate the $G(Q)$ factor of standard WI models. To validate the use of the above analytical form the $G(Q)$ factor in our framework, we generate the numerical $G(Q)$ factor for MWI model with the three $\alpha$-attractor potentials using the WI2easy code \cite{Rodrigues:2025neh} and compared it with the above analytical form. Fig.~\ref{GQ-compare} displays the numerical forms of the $G(Q)$ factor for three different $\alpha$-attractor potentials as well as the above analytical form and confirms that within the working range of $Q$, the analytical form of the $G(Q)$ is in excellent agreement with the numerically generated ones using the WI2easy code. 

Moreover, to determine the scalar spectral index in this scenario, we ignore the thermalization of the inflaton field in strong dissipative regime as it doesn't effect the scalar power spectrum much in this regime. The analytic form of the $n_s$ was first determined in \cite{Das:2022ubr} and can be given as 
\begin{eqnarray}
n_s\approx 1- \frac{(14-10{\mathcal A}(Q))\epsilon_V+6{\mathcal A}(Q)\eta_V}{7Q},
\label{ns-MWI}
\end{eqnarray}
where we have defined ${\mathcal A}(Q)$ as 
\begin{eqnarray}
{\mathcal A}(Q)=\frac{3+2\pi Q}{3+4\pi Q}+Q\frac{d\ln G(Q)}{dQ}.
\end{eqnarray}
With this form of $n_s$ and defining $r$ as given in Eq.~(\ref{r-WI}), we can now analyze the attractor behaviour in WI in the $n_s-r$ plane. However, before that, we will try to estimate whether WI gracefully exits in such $\alpha$-attractor models and whether graceful exit places any constraint on the model parameters.

\section{Analysis of Graceful Exit in Warm $\alpha$-attractors}
\label{graceful-exit}

It was pointed out in \cite{Das:2020lut} that, unlike in CI, the graceful exit conditions are more intricate in WI. In CI, a potential yielding increasing $\epsilon_V$ is sufficient for graceful exit. However, WI takes place when $\epsilon_V\ll1+Q$ and gracefully exits when $\epsilon_V\sim 1+Q$. In WI both $\epsilon_V$ and $Q$ evolve with $e$-folds. Evolution of $\epsilon_V$ depends solely on the form of the potential, whereas the evolution of $Q$ depends on both the potential as well as the form of the dissipative coefficient. The evolution of $\epsilon_V$ and $Q$ w.r.t. the $e$-foldings $N$ can be written as \cite{Das:2020lut}
\begin{eqnarray}
\frac{d\ln\epsilon_V}{dN}&=&\frac{4\epsilon_V-2\eta_V}{1+Q}, \label{eps-evo} \\
\frac{d\ln Q}{dN}&=&\frac{(2p+4)\epsilon_V-2p\eta_V-4c\kappa_V}{C_Q} \label{Q-evo},
\end{eqnarray}
where $C_Q\equiv 4-p+(4+p)Q$. Therefore, there are five distinct ways in which WI can gracefully exits:
\begin{enumerate}
\item {\bf When $\epsilon_V$ is constant:} In such a case, graceful exit can only occur if $Q$ decreases.
\item {\bf When $\epsilon_V$ increases:} In such a case, graceful exit can occur in three different ways
\begin{enumerate}
\item When $Q$ remains constant
\item When $Q$ decreases
\item When $Q$ increases but with a slower rate than that of $\epsilon_V$
\end{enumerate}
\item {\bf When $\epsilon_V$ decreases:} In such a case, graceful exit can only occur if $Q$ decreases faster than $\epsilon_V$.
\end{enumerate}

We will now analyse how WI ends in these three types of attractor models and whether graceful exit conditions put any bound on the parameters in the inflaton potential. As we have chosen the MWI model for our analysis, we will consider $p=3$ and $c=0$ in the form of the dissipative coefficient (given in Eq.~(\ref{dis-coeff})) from now onwards. 

\subsection{Graceful exit analysis for T-model}

For the T-model, we note from Eq.~(\ref{eps-evo}) that 
\begin{eqnarray}
\frac{d\ln\epsilon_V}{dN}=\frac{1}{1+Q}\left(\frac{8n\cosh\left(\sqrt{2/3\alpha}\phi\right)}{3\alpha\sinh^2\left(\sqrt{2/3\alpha}\phi\right)}\right),
\end{eqnarray}
which is always positive indicating that $\epsilon_V$ is always increasing in this scenario.  Thus graceful exit is always ensured when $Q$ decreases or remains constant. Moreover, if $Q$ increases but with a slower rate than that of $\epsilon_V$, then also graceful exit will take place. 

From Eq.~(\ref{Q-evo}) we get that 
\begin{eqnarray}
\frac{d\ln Q}{dN}=\frac{1}{1+7Q}\left(\frac{-8n^2+24n\cosh\left(\sqrt{2/3\alpha}\phi\right)}{3\alpha\sinh^2\left(\sqrt{2/3\alpha}\phi\right)}\right)
\end{eqnarray}
Therefore, $Q$ decreases when $n>3\cosh(\sqrt{2/3\alpha}\phi)$, remains constant when $n=3\cosh(\sqrt{2/3\alpha}\phi)$, and increases when $n<3\cosh(\sqrt{2/3\alpha}\phi)$. Thus, it is easy to see that graceful exit is ensured when $n\geq3\cosh(\sqrt{2/3\alpha}\phi)$. 

To attain graceful exit when $Q$ increases we need to demand 
\begin{eqnarray}
\frac{d\ln\epsilon_V}{dN}>\frac{d\ln Q}{dN},
\end{eqnarray}
which for strong dissipation ($Q\gg1$) yields $n>-4\cosh(\sqrt{2/3\alpha}\phi)$. However, as $n$ is a positive number, we get the bound on $n$ when $Q$ is increasing as $0<n<3\cosh(\sqrt{2/3\alpha}\phi)$. Hence, combining both these bounds on $n$ ($0<n<3\cosh(\sqrt{2/3\alpha}\phi)$ when $Q$ increases and $n\geq3\cosh(\sqrt{2/3\alpha}\phi)$ when $Q$ decreases or remains constant), we see that graceful exit in T-model is always ensured for all values of $n$ irrespective of the values of $\alpha$.

\subsection{Graceful exit analysis for E-model}

The graceful exit situation for E-model is similar to the one in T-model. In E-model the evolutions of $\epsilon_V$ and $Q$ are given as 
\begin{eqnarray}
\frac{d\ln\epsilon_V}{dN}&=&\frac{1}{1+Q}\left(\frac{8ne^{\sqrt{2/3\alpha}\phi}}{3\alpha(e^{\sqrt{2/3\alpha}\phi}-1)^2}\right),\\
\frac{d\ln Q}{dN}&=&\frac{1}{1+7Q}\left(\frac{-8n^2+24ne^{\sqrt{2/3\alpha}\phi}}{3\alpha(e^{\sqrt{2/3\alpha}\phi}-1)^2}\right).
\end{eqnarray}
It is easy to see that $\epsilon_V$ is always increasing, and $Q$ decreases or remains constant when $n\geq 3e^{\sqrt{2/3\alpha}\phi}$. On the other hand, $Q$ increases but slower than $\epsilon_V$ puts a constraint on $n$ as $0<n<3e^{\sqrt{2/3\alpha}\phi}$. Therefore, graceful exits takes place for all values of $n$ without any dependence on $\alpha$. 

\subsection{Graceful exit analysis for Polynomial model}

As per the discussion of the polynomial model in CI, we will be interested in $\phi>\mu$ as well as $n\gg1$. In this limit it is legitimate to drop the terms $(\mu/\phi)^{2n}$ in comparison to unity in the expressions of $\epsilon_V$ and $\eta_V$ given in Eqs.~(\ref{slow-roll-param-poly}) which will simplify our discussion. We can write the slow-roll parameters with this approximation as 
\begin{eqnarray}
\epsilon_V\sim\frac{2n^2}{\mu^2}\left(\frac\mu\phi\right)^{4n+2},\quad\quad\quad\quad \eta_V\sim -\frac{2n(n+1)}{\mu^2}\left(\frac\mu\phi\right)^{2n+2}.
\end{eqnarray} 
As $\eta_V$ is negative in this model (and $\epsilon_V$ is always positive) we can infer from Eq.~(\ref{eps-evo} and Eq.~(\ref{Q-evo})) that both $\epsilon_V$ and $Q$ are increasing in this scenario. And to let $Q$ increase slower than $\epsilon_V$ we need $(\phi/\mu)^{2n}>-(9/4)(n/1+n)$, which is always true. Thus graceful exit will always occur in the Polynomial model irrespective of the values of $n$ and $\mu$ (or $\alpha$). 

\section{Analysis of the behaviour of the $\alpha$-attractor models in Warm Inflation}
\label{attractor-WI}

From the analysis of the attractor behaviour of the $\alpha$-attractor models in CI as presented in Sec.~\ref{attractor-CI} we noted that $n_s$ and $r$ are required to be expressed in terms of $N_*$. In MWI, $n_s$ depends on $\phi$ (through the slow-roll parameters) as well as $Q$, as can be seen from Eq.~(\ref{ns-MWI}). The value of the inflaton field at the pivot-scale crossing, $\phi_*$, can be determined in terms of $N_*$ as \cite{Das:2019acf}
\begin{eqnarray}
N_*=-\frac{1}{M_{\rm Pl}^2}\int_{\phi_*}^{\phi_e}d\phi\frac{V}{V_{,\phi}}(1+Q)\sim -\frac{1}{M_{\rm Pl}^2}\int_{\phi_*}^{\phi_e}d\phi\frac{V}{V_{,\phi}}Q,
\end{eqnarray}
which is similar to CI (as given in Eq.~(\ref{Nstar-CI})). In WI, $\phi_e$, the value of the inflaton field at the end of inflation, can be calculated  by setting $\epsilon_V\sim1+Q$, which is the condition of end of WI. Moreover, as $Q\sim T^3/3H$ in MWI, and as the $\alpha$-attractor dynamics always takes place in the plateau regions of the $\alpha$-attractor potentials, approximating $H\sim V_0/3M_{\rm Pl}^2$, we note that in MWI with $\alpha$-attractor potentials $Q$ doesn't depend on $\phi$. Therefore, we can further simplify the above equation of $N_*$ as 
\begin{eqnarray}
N_*\sim -\frac{Q}{M_{\rm Pl}^2}\int_{\phi_*}^{\phi_e}d\phi\frac{V}{V_{,\phi}}.
\end{eqnarray}
This equation will help us determine $\phi_*$ in terms of $N_*$. Furthermore, in slow-roll regime of WI, it is easy to see from the slow-roll approximated background dynamics of WI, that one can express $Q$ in terms of $\phi$ as \cite{Das:2020lut}
\begin{eqnarray}
(1+Q)^{2p}Q^{4-p}=\frac{M_{\rm Pl}^{2p+4}M^{4(1-p-c)}C_\Upsilon^4}{2^{2p}3^2C_r^p}\frac{V_{,\phi}^{2p}}{V^{p+2}}\phi^{4c},
\end{eqnarray}
and for strongly dissipative MWI model the above equation yields \cite{Das:2019acf}
\begin{eqnarray}
Q^7\sim \frac{M_{\rm Pl}^{10}M^{-8}C_\Upsilon^4}{2^63^2C_r^3}\frac{V_{,\phi}^{6}}{V^{5}}.
\end{eqnarray}
Thus, knowing $\phi_*$ in terms of $N_*$, one gets $Q_*$ (the value of $Q$ at the pivot-scale crossing) in terms of $N_*$. Hence, incorporating this method, one can obtain $n_s$ in terms of $N_*$ as we did for CI in Sec.~\ref{attractor-CI}.

To obtain $r$ in terms of $N_*$ we need to further quantify the factor $T/H$ appearing in the scalar power spectrum of WI as given in Eq.~(\ref{scalar-ps}). Again, it is straightforward to show that in the slow-roll regime of strong dissipative WI one gets \cite{Das:2019acf}
\begin{eqnarray}
\frac TH\sim\frac{3^{3/4}M_{\rm Pl}^{3/2}}{(12C_r)^{1/4}}\frac{V_{,\phi}^{1/2}}{V^{3/4}}Q^{-1/4},
\end{eqnarray}
which indicates that $T/H$ depends on $\phi$ as well as on $Q$. As we have already devised a method to express $\phi_*$ and $Q_*$ in terms of $N_*$, $T/H$ can also be expressed in terms of $N_*$ using the above expression. 

However, it is important to note that both $n_s$ and $r$ in WI depends on the model parameters $C_{\Upsilon}$, $M$, $C_r$ and $V_0$, unlike the case in CI. Even if we fix $C_{\Upsilon}$, $M$ and $C_r$ for all the models while varying the parameter $n$ and $\alpha$ of the $\alpha$-attractor potentials, one needs to change $V_0$ in order to match the scalar amplitude at the pivot-scale, $A_s\sim 2.09\times 10^{-9}$, with the observed data. Hence, in the $n_s-r$ plot in WI the various points will belong to various $V_0$ as we will see below. This was not the case when we previously analyzed the $n_s-r$ plots of $\alpha-$attractors in CI in Sec.~\ref{attractor-CI}. 


\subsection{Behaviour of T-model in Warm Inflation}

\begin{center}
\begin{figure*}[!htb]
\subfigure[$n=1$]{\includegraphics[width=5.7cm]{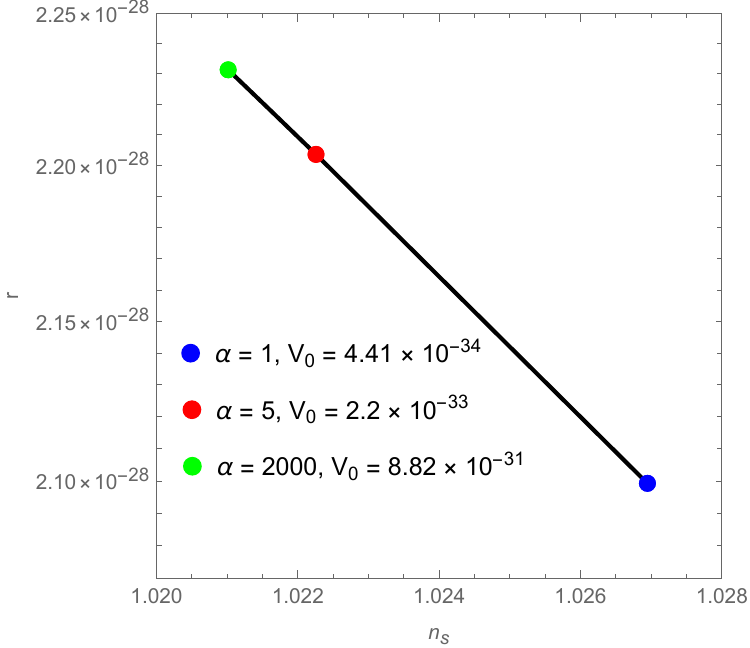}}
\subfigure[$n=2$]{\includegraphics[width=5.7cm]{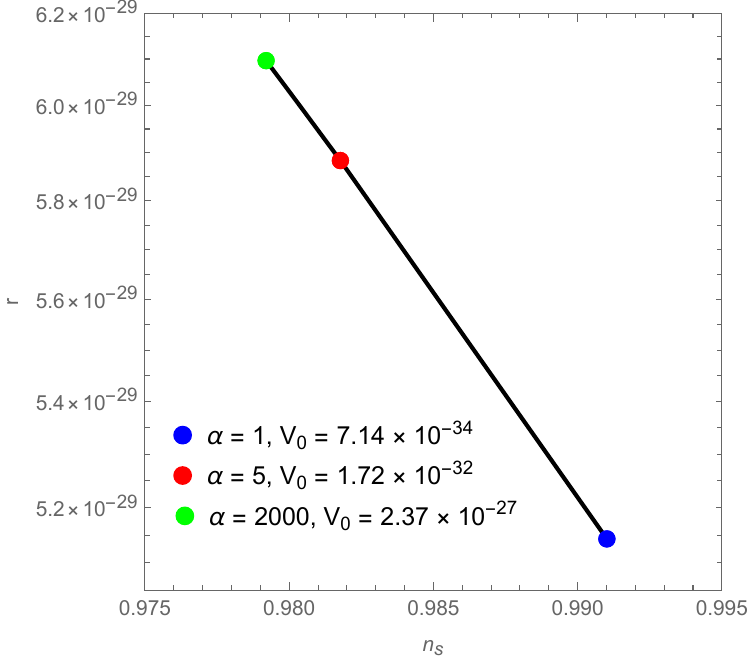}}
\subfigure[$n=3$]{\includegraphics[width=5.7cm]{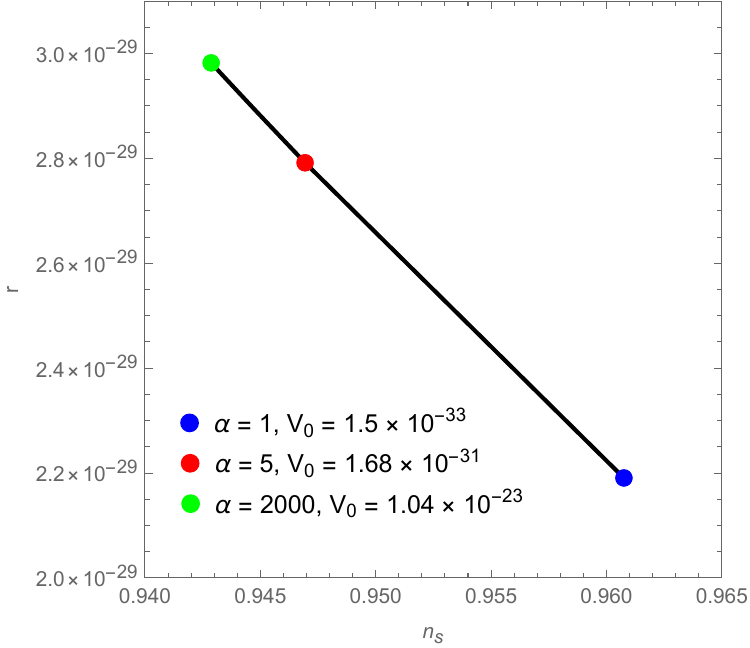}}
\caption{The behaviour of $n_s$ vs $r$ plot of T-model in Warm inflation as a function of the parameter $\alpha$. We have chosen the models parameters as $C_\Upsilon=5\times10^{-11}$, $M=3.93\times10^{-13}M_{\rm Pl}$ and $g_*=106.75$. The values of $V_0$ quoted in the plots are in units of $M_{\rm Pl}^4$. These plots are shown for $N_*=51$.}
\label{T-model-WI-1}
\end{figure*}
\end{center}

\begin{center}
\begin{figure}[!htb]
\includegraphics[width=15cm]{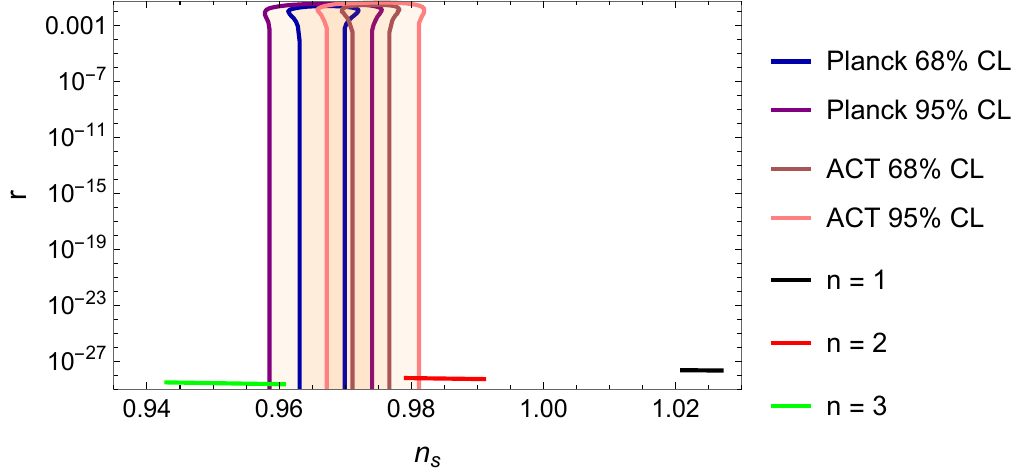}
\caption{The $n_s$ vs $r$ plot for T-model in WI for $N_*=51$}
\label{T-model-nsr-WI}
\end{figure}
\end{center}

Using the T-model potential given in Eq.~(\ref{T-alpha}) it is easy to derive, following the methods prescribed above, the values of $\phi_e$ and $\phi_*$ as 
\begin{eqnarray}
\cosh\left(\sqrt{\frac{2}{3\alpha}}\phi_e\right)&=&\sqrt{\frac{3\alpha(1+Q_e)+4n^2}{3\alpha(1+Q_e)}},\\
\cosh\left(\sqrt{\frac{2}{3\alpha}}\phi_*\right)
&=&\frac{4n}{3\alpha(1+Q_*)}N_*+\sqrt{\frac{3\alpha(1+Q_e)+4n^2}{3\alpha(1+Q_e)}}.
\end{eqnarray}
Here, $Q_*$ and $Q_e$ designates the values of $Q$ at the pivot-scale crossing and at the end of WI, respectively. However, we will consider $Q_e\sim Q_*$ as $Q$ doesn't vary significantly throughout the evolution as has been argued above. 
Once we have $\phi_*$, it is then straightforward to calculate $n_s$ and $r$ in terms of $N_*$, though the analytic forms of $n_s$ and $r$ are not as simple as in CI. We then plot $n_s$ vs. $r$ as a function of $\alpha$ for $n=1,\,2$ and 3, which we depict in Fig.~\ref{T-model-WI-1}. If we plot these models together in the $n_s-r$ plane, as has been shown in Fig.~\ref{T-model-nsr-WI}, we see that the attractor behaviour that was there in CI is now lost in a WI setup. 

It is important to note that it might seem from Fig.~\ref{T-model-nsr-WI} that the T-model is incompatible with the current data, especially the $n=1$ case. However, the reason of such an appearance is due to the chosen values of the model parameters. It might happen that for a different set of model parameters $n=1$ can be made compatible with the data, which will leave other $n$ values out of the range of the observed contours in the $n_s-r$ plane. The main point to highlight here is that the choice of the model parameters here is only representative, and any other set of choice may shift these plots. However, even then the absence of the attractor behaviour will not change. Here, our primary goal is to check the attractor behaviour in the $n_s-r$ plane, and not to fit the models with the data. We leave such an analysis for a future work. 


\subsection{Behaviour of E-model in Warm Inflation}

\begin{center}
\begin{figure*}[!htb]
\subfigure[$n=1$]{\includegraphics[width=5.7cm]{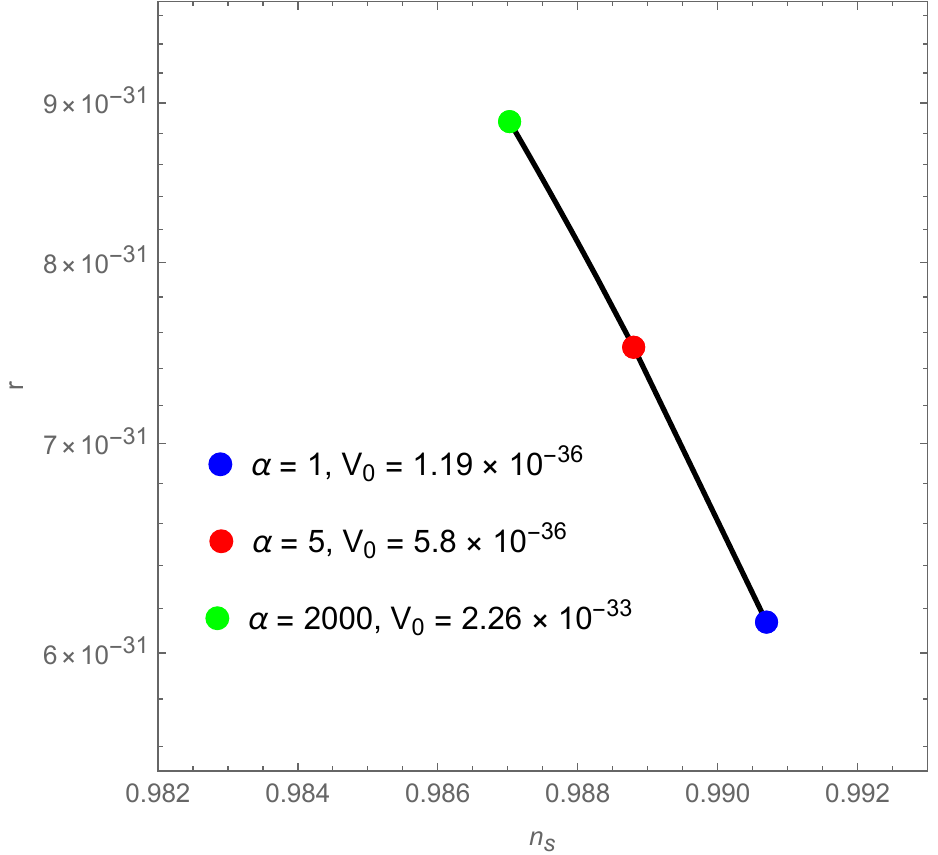}}
\subfigure[$n=2$]{\includegraphics[width=5.7cm]{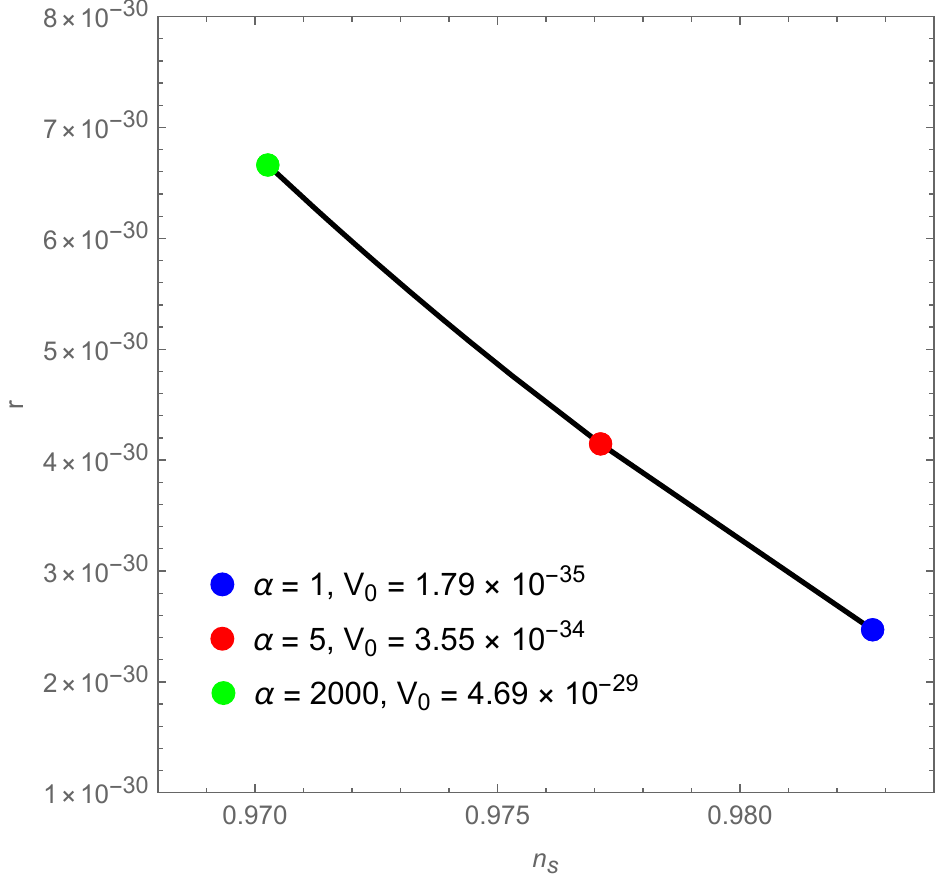}}
\subfigure[$n=3$]{\includegraphics[width=5.7cm]{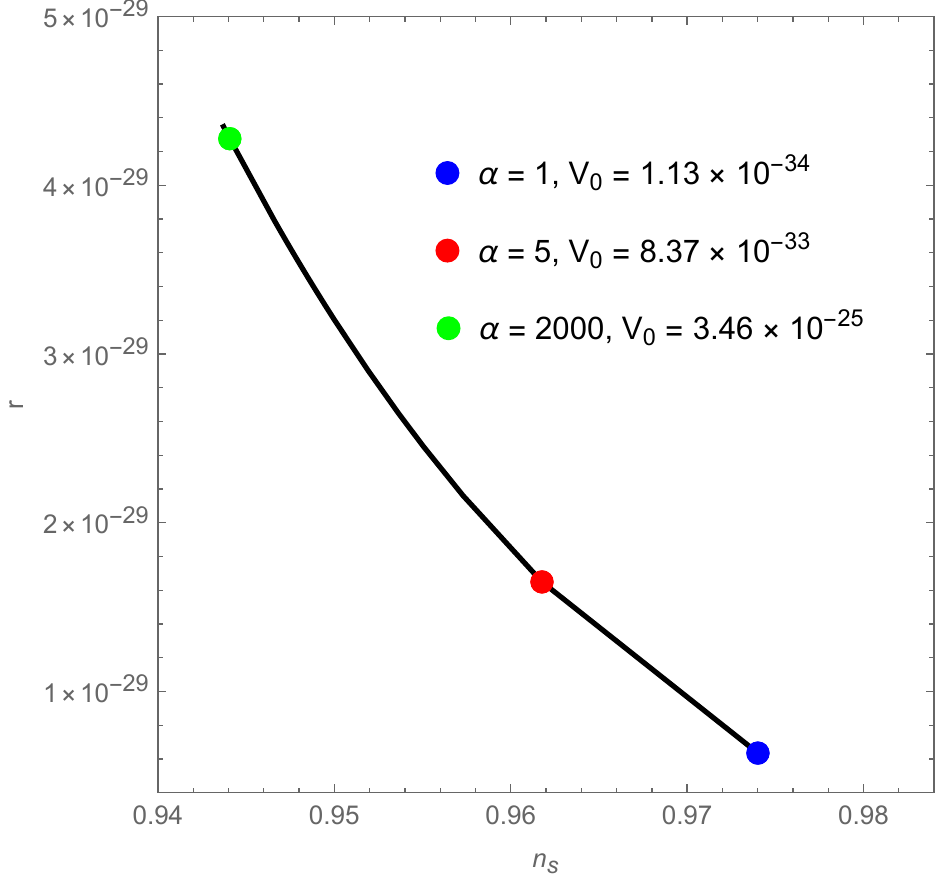}}
\caption{The behaviour of $n_s$ vs $r$ plot of E-model in Warm inflation as a function of the parameter $\alpha$. We have chosen the models parameters as $C_\Upsilon=5\times10^{-11}$, $M=3.93\times10^{-13}M_{\rm Pl}$ and $g_*=106.75$. The values of $V_0$ quoted in the plots are in units of $M_{\rm Pl}^4$. These plots are shown for $N_*=51$.}
\label{E-model-WI-1}
\end{figure*}
\end{center}

\begin{center}
\begin{figure}[!htb]
\includegraphics[width=15cm]{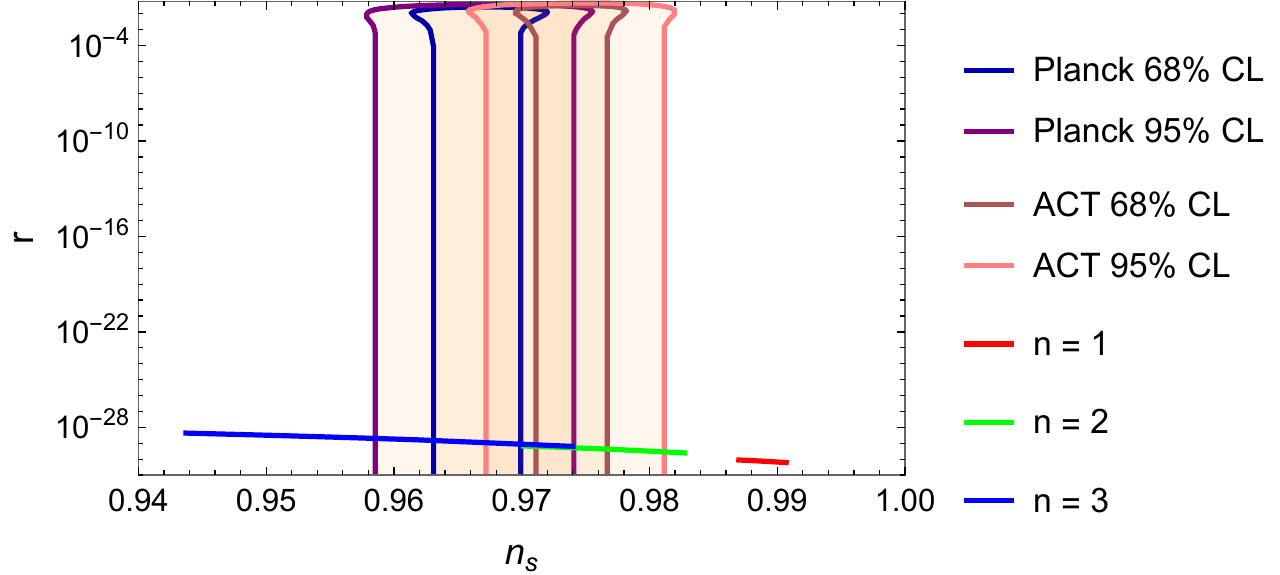}
\caption{The $n_s$ vs $r$ plot for E-model in WI for $N_*=51$}
\label{E-model-nsr-WI}
\end{figure}
\end{center}

In a similar way, one can calculate $\phi_e$ and $\phi_*$ for the E-model potential (Eq.~(\ref{E-alpha})) in WI as 
\begin{eqnarray}
 e^{\sqrt{\frac{2}{3\alpha}}\phi_e}&=&\left(1+\frac{2n}{\sqrt{3\alpha(1+Q_e)}}\right),\\
 e^{\sqrt{\frac{2}{3\alpha}}\phi_*}&=&-\frac{4n N_{\star}}{3\alpha(1+Q_*)}-\left(1+\frac{2n}{\sqrt{3\alpha(1+Q_e)}}\right)+\log\left[1+\frac{2n}{\sqrt{3\alpha(1+Q_*)}}\right]\nonumber\\
 &&-W_{-1}\left(-e^{-1-\frac{4n N_{\star}}{3\alpha(1+Q_*)}-\frac{2n}{\sqrt{3\alpha(1+Q_*)}}+\log\left[1+\frac{2n}{\sqrt{3\alpha(1+Q_*)}}\right]}\right),
\end{eqnarray}
and here, too, we will consider $Q_e\sim Q_*$. We display the nature of the $n_s$ vs. $r$ plot as a function of $\alpha$ for such a model in Fig.~\ref{E-model-WI-1} for $n=1,\,2$ and 3, and Fig.~\ref{E-model-nsr-WI} shows that the CI attractor behaviour for E-model is no longer there in WI. Moreover, as we discussed in the case of T-model, here, too, the plots are given for some representative set of model parameters, and each of such models may comply with the data for some different set of parameter values. Yet, the attractor behaviour shown in Fig.~\ref{E-model-nsr-WI} will not change. 

\subsection{Behaviour of Polynomial model in Warm Inflation}

\begin{center}
\begin{figure*}[!htb]
\subfigure[$\alpha=1$]{\includegraphics[width=5.7cm]{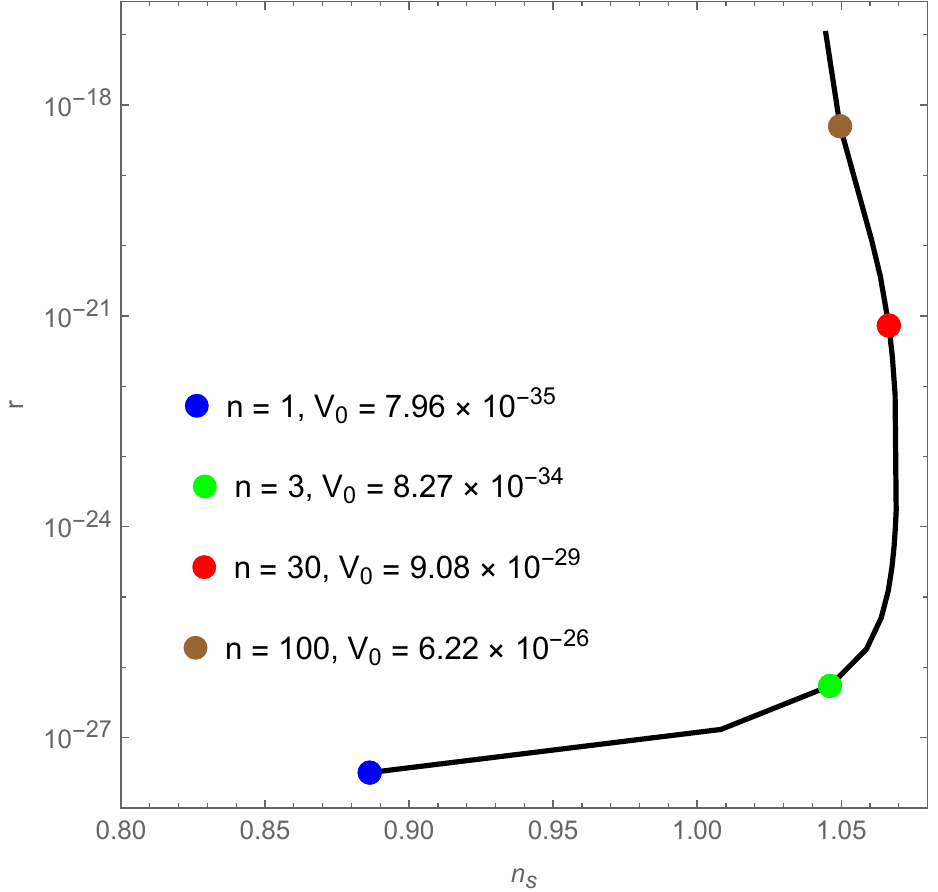}}
\subfigure[$\alpha=50$]{\includegraphics[width=5.7cm]{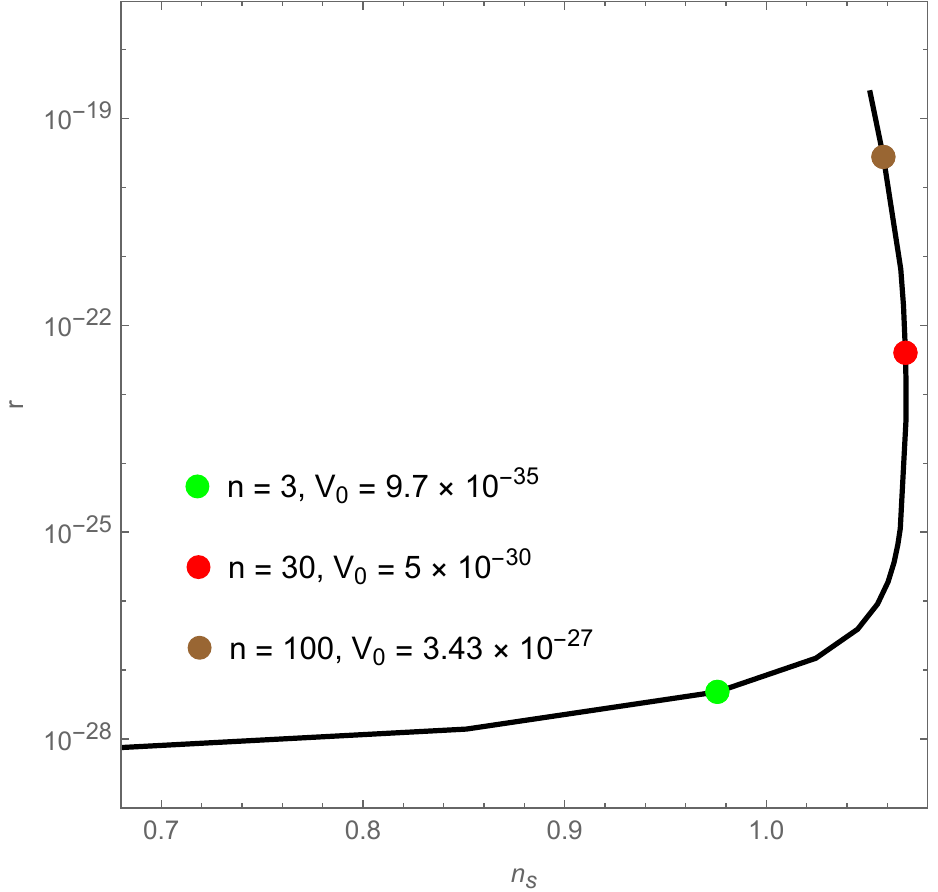}}
\caption{The behaviour of $n_s$ vs $r$ plot of Polynomial model in Warm inflation as a function of the parameter $n$. We have chosen the models parameters as $C_\Upsilon=5\times10^{-11}$, $M=3.93\times10^{-13}M_{\rm Pl}$ and $g_*=106.75$. The values of $V_0$ quoted in the plots are in units of $M_{\rm Pl}^4$. These plots are shown for $N_*=51$.}
\label{Pol-model-WI-1}
\end{figure*}
\end{center}

\begin{center}
\begin{figure}[!htb]
\includegraphics[width=15cm]{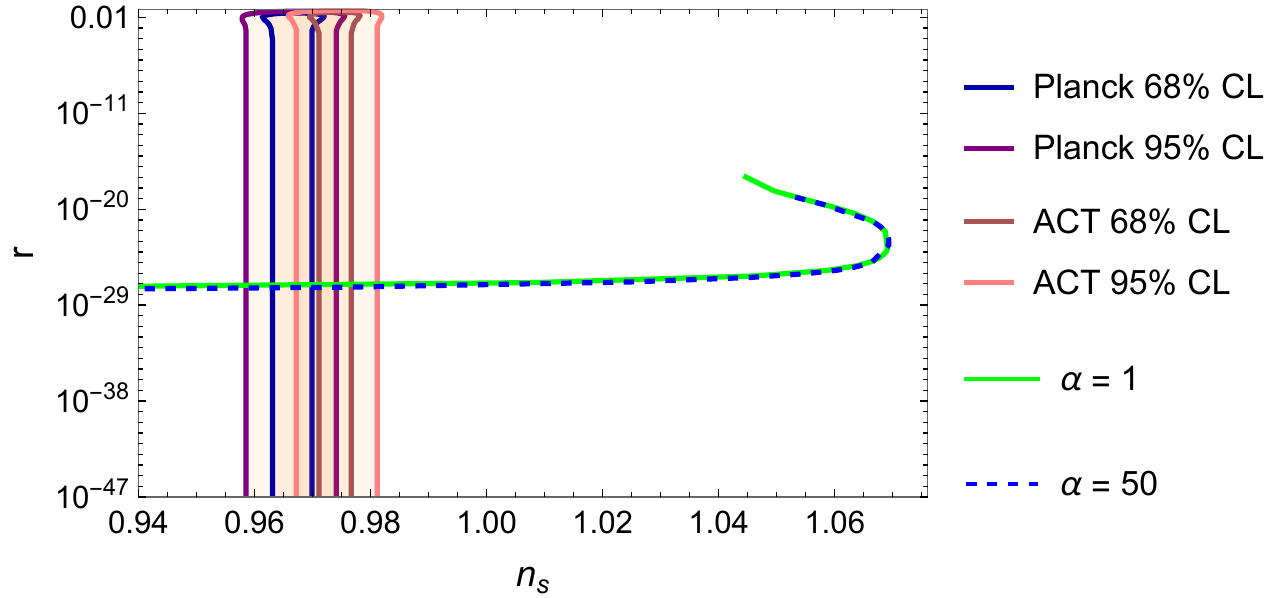}
\caption{The $n_s$ vs $r$ plot for Polynomial model in WI for $N_*=51$}
\label{Pol-model-nsr-WI}
\end{figure}
\end{center}

We derive $\phi_e$ and $\phi_*$ for Polynomial model potential in WI as 
\begin{eqnarray}
 \frac{\phi_e}{\mu}&=&2^{\frac{1}{2+4n}}\left(\frac{\mu^2(1+Q_e)}{n^2}\right)^{-\frac{1}{2+4n}},\\
\frac{\phi_*}{\mu}&=&\left(\frac{2}{\mu}\right)^{\frac1{1+n}}\left(\frac{n(n+1)N_*}{1+Q_*}+\frac{2^{\frac{1+n}{1+2n}}n^{\frac{2+2n}{1+2n}}\mu^{\frac{2n}{1+2n}}}{(1+Q_e)^{\frac{1+n}{1+2n}}}\right)^{\frac1{2+2n}},
\end{eqnarray}
and like previous cases, we will consider $Q_e\sim Q_*$. It is to note that in CI, the attractor behaviour of the Polynomial model was observed in the parameter $n$. Hence, in WI too, we would like to see the $n_s-r$ behaviour as a function of $n$. We portray the $n_s$ vs $r$ plot for Polynomial model as a function of $n$ for fixed values of $\alpha$ ($\alpha=1$ and $\alpha=50$) in Fig.~(\ref{Pol-model-WI-1}). We noticed that the behaviour, unlike in the CI case, depends very weakly on the parameter $\alpha$. When plotted together, as in Fig.~(\ref{Pol-model-nsr-WI}), we see that the attractor behaviour of the Polynomial model is also lost in a Warm Inflation setup. Once again, we would like to emphasise that the choice of model parameters are only representative, as in the previous two cases.

\section{Discussion and Conclusion}
\label{conclusion}

The novel property of the T- and E-type $\alpha$-attractor models in a CI setup is that the behaviour of the $n_s$ vs $r$ graphs in $n_s-r$ plane show an attractor behaviour for various values of $n$ when the model parameter $\alpha$ tends to zero  \cite{Kallosh:2013yoa, Kallosh:2013hoa}. On the other hand, the Polynomial models show similar attractor behaviour in the $n_s-r$ plane for various values of $\alpha$ when the potential parameter $n$ tends to zero \cite{Kallosh:2022feu}. Such an attractor behaviour puts these models in the sweet-spot of the $n_s-r$ plot of the Planck observations. 

In this article we analyse the behaviour of these attractor models in a WI setup. We have chosen Minimal Warm Inflation model \cite{Berghaus:2019whh} for our analysis where WI takes place in the strong dissipative regime. In such a model the dissipative coefficient is a function of $T$ alone and has no dependence on $\phi$. Such features of MWI helps ease the analysis of determining the $n_s-r$ behaviour of the attractor models, as discussed in Sec.~\ref{WI-summary}. Moreover, as in the strong dissipative regime WI is maximally away from the CI dynamics, the effect on $n_s-r$ plot due to WI dynamics is expected to be  the most in this regime. We showed in Sec.~\ref{graceful-exit}  that MWI model gracefully exits in these type of attractor models, and the graceful exit conditions don't put any constraint on the model parameters. Our final results were furnished in Sec.~\ref{attractor-WI} where we observed that in all these attractor models the attractor behaviour, which is the main feature of such models in CI, is lost in strongly dissipative MWI model. This strongly indicates that the attractor nature of $\alpha$-attractor models is unique to CI alone, and any departure from the CI dynamics will lead to different behaviours of those models in the $n_s-r$ plane. However, on a positive note, the $n_s-r$ behaviour in the $n_s-r$ plots in WI obtained through the analysis done here, indicates that within specific parameter ranges, such attractor models can now be made compatible with the recent ACT results in a WI setup. As a passing comment, we would like to mention that upcoming CMB observations, like {\it Lite}BIRD \cite{LiteBIRD:2022cnt} and the Simons Observatory \cite{SimonsObservatory:2018koc}, aim to detect the B-mode polarization of the CMB which can put to test many viable inflationary models, cold or warm. While the Simons Observatory aims to detect the B-mode signal at $r\sim 0.01$ \cite{SimonsObservatory:2018koc}, a non-detection of B-modes in {\it Lite}BIRD will put an upper bound on the tensor-to-scalar ratio as $r<0.002$ \cite{LiteBIRD:2022cnt}. These observations can potentially rule out many CI scenarios producing large tensor-to-scalar ratio. However, as the WI models in strong dissipation, like the ones studied here, generally produce smaller $r$ ($r<10^{-6}$), these models can potentially be ruled out if any of these future detections detects a B-mode of the order $10^{-2}$ or even $10^{-3}$. Above all, as discussed before, WI in strong dissipative regime, as is the case in MWI studied here, can give rise to bispectrum non-Gaussianities with distinct shapes (``new-warm'' \cite{Mirbabayi:2022cbt}) that do not arise in CI. Thus, a detection of such shapes in the primordial non-Gaussian signatures can be a smoking gun test for the WI models studied here.

It is of importance to note that the choices of model parameters made in the paper are all representative. A detailed analysis of best-fit model parameters of these attractor models in WI is of importance, as our plots clearly hints that in WI, these models can be made in tune with the recent ACT results \cite{ACT:2025fju, ACT:2025tim} within certain parameter ranges, which is not the case in a CI setup anymore. A methodology to analyse WI models given observational data has been recently developed in \cite{Kumar:2024hju}. Following this methodology we will like to explore these attractor models in WI in future which will put constraint on the parameters of these models. Moreover, it is of importance to analyse the attractor behaviour of these models in weak dissipative WI models which are closer to the dynamics of CI. It might happen that in weak dissipative regime, as the WI dynamics is closer to CI dynamics, these models retain their attractor behaviour even in a WI setup. We leave such an analysis for a future work. 

{\it Special note:} While this paper was about to be submitted in arXiv, another paper appeared on arXiv \cite{Saha:2025yzf} where the authors have analysed $\alpha$-attractor E-model in a WI setup (both in weak and strong dissipative regime) where the dissipative coefficient is proportional to $T$. The E-model was tested against the Planck-2018 results and it was shown that the E-model in WI is compatible with these data, whereas the attractor behaviour of E-model is not analysed in this paper. 

\acknowledgements

The work of S.D. is supported by the Start-up Research Grant (SRG) awarded by Anusandhan
National Research Foundation (ANRF), Department of Science and Technology, Government
of India [File No. SRG/2023/000101/PMS]. The work of D.C. at Ashoka University was supported by the postdoctoral fellowship provided by Axis Bank. D.C. warmly acknowledges Axis Bank for this fellowship. 



\label{Bibliography}
\bibliography{refs}

\begin{thebibliography}{58}%
\makeatletter
\providecommand \@ifxundefined [1]{%
 \@ifx{#1\undefined}
}%
\providecommand \@ifnum [1]{%
 \ifnum #1\expandafter \@firstoftwo
 \else \expandafter \@secondoftwo
 \fi
}%
\providecommand \@ifx [1]{%
 \ifx #1\expandafter \@firstoftwo
 \else \expandafter \@secondoftwo
 \fi
}%
\providecommand \natexlab [1]{#1}%
\providecommand \enquote  [1]{``#1''}%
\providecommand \bibnamefont  [1]{#1}%
\providecommand \bibfnamefont [1]{#1}%
\providecommand \citenamefont [1]{#1}%
\providecommand \href@noop [0]{\@secondoftwo}%
\providecommand \href [0]{\begingroup \@sanitize@url \@href}%
\providecommand \@href[1]{\@@startlink{#1}\@@href}%
\providecommand \@@href[1]{\endgroup#1\@@endlink}%
\providecommand \@sanitize@url [0]{\catcode `\\12\catcode `\$12\catcode
  `\&12\catcode `\#12\catcode `\^12\catcode `\_12\catcode `\%12\relax}%
\providecommand \@@startlink[1]{}%
\providecommand \@@endlink[0]{}%
\providecommand \url  [0]{\begingroup\@sanitize@url \@url }%
\providecommand \@url [1]{\endgroup\@href {#1}{\urlprefix }}%
\providecommand \urlprefix  [0]{URL }%
\providecommand \Eprint [0]{\href }%
\providecommand \doibase [0]{http://dx.doi.org/}%
\providecommand \selectlanguage [0]{\@gobble}%
\providecommand \bibinfo  [0]{\@secondoftwo}%
\providecommand \bibfield  [0]{\@secondoftwo}%
\providecommand \translation [1]{[#1]}%
\providecommand \BibitemOpen [0]{}%
\providecommand \bibitemStop [0]{}%
\providecommand \bibitemNoStop [0]{.\EOS\space}%
\providecommand \EOS [0]{\spacefactor3000\relax}%
\providecommand \BibitemShut  [1]{\csname bibitem#1\endcsname}%
\let\auto@bib@innerbib\@empty
\bibitem [{\citenamefont {Aghanim}\ \emph {et~al.}(2020)\citenamefont {Aghanim}
  \emph {et~al.}}]{Planck:2018vyg}%
  \BibitemOpen
  \bibfield  {author} {\bibinfo {author} {\bibfnamefont {N.}~\bibnamefont
  {Aghanim}} \emph {et~al.} (\bibinfo {collaboration} {Planck}),\ }\href
  {\doibase 10.1051/0004-6361/201833910} {\bibfield  {journal} {\bibinfo
  {journal} {Astron. Astrophys.}\ }\textbf {\bibinfo {volume} {641}},\ \bibinfo
  {pages} {A6} (\bibinfo {year} {2020})},\ \bibinfo {note} {[Erratum:
  Astron.Astrophys. 652, C4 (2021)]},\ \Eprint
  {http://arxiv.org/abs/1807.06209} {arXiv:1807.06209 [astro-ph.CO]}
  \BibitemShut {NoStop}%
\bibitem [{\citenamefont {Ade}\ \emph {et~al.}(2021)\citenamefont {Ade} \emph
  {et~al.}}]{BICEP:2021xfz}%
  \BibitemOpen
  \bibfield  {author} {\bibinfo {author} {\bibfnamefont {P.~A.~R.}\
  \bibnamefont {Ade}} \emph {et~al.} (\bibinfo {collaboration} {BICEP, Keck}),\
  }\href {\doibase 10.1103/PhysRevLett.127.151301} {\bibfield  {journal}
  {\bibinfo  {journal} {Phys. Rev. Lett.}\ }\textbf {\bibinfo {volume} {127}},\
  \bibinfo {pages} {151301} (\bibinfo {year} {2021})},\ \Eprint
  {http://arxiv.org/abs/2110.00483} {arXiv:2110.00483 [astro-ph.CO]}
  \BibitemShut {NoStop}%
\bibitem [{\citenamefont {Kallosh}\ \emph {et~al.}(2013)\citenamefont
  {Kallosh}, \citenamefont {Linde},\ and\ \citenamefont
  {Roest}}]{Kallosh:2013yoa}%
  \BibitemOpen
  \bibfield  {author} {\bibinfo {author} {\bibfnamefont {R.}~\bibnamefont
  {Kallosh}}, \bibinfo {author} {\bibfnamefont {A.}~\bibnamefont {Linde}}, \
  and\ \bibinfo {author} {\bibfnamefont {D.}~\bibnamefont {Roest}},\ }\href
  {\doibase 10.1007/JHEP11(2013)198} {\bibfield  {journal} {\bibinfo  {journal}
  {JHEP}\ }\textbf {\bibinfo {volume} {11}},\ \bibinfo {pages} {198} (\bibinfo
  {year} {2013})},\ \Eprint {http://arxiv.org/abs/1311.0472} {arXiv:1311.0472
  [hep-th]} \BibitemShut {NoStop}%
\bibitem [{\citenamefont {Kallosh}\ and\ \citenamefont
  {Linde}(2013{\natexlab{a}})}]{Kallosh:2013hoa}%
  \BibitemOpen
  \bibfield  {author} {\bibinfo {author} {\bibfnamefont {R.}~\bibnamefont
  {Kallosh}}\ and\ \bibinfo {author} {\bibfnamefont {A.}~\bibnamefont
  {Linde}},\ }\href {\doibase 10.1088/1475-7516/2013/07/002} {\bibfield
  {journal} {\bibinfo  {journal} {JCAP}\ }\textbf {\bibinfo {volume} {07}},\
  \bibinfo {pages} {002} (\bibinfo {year} {2013}{\natexlab{a}})},\ \Eprint
  {http://arxiv.org/abs/1306.5220} {arXiv:1306.5220 [hep-th]} \BibitemShut
  {NoStop}%
\bibitem [{\citenamefont {Kallosh}\ and\ \citenamefont
  {Linde}(2022)}]{Kallosh:2022feu}%
  \BibitemOpen
  \bibfield  {author} {\bibinfo {author} {\bibfnamefont {R.}~\bibnamefont
  {Kallosh}}\ and\ \bibinfo {author} {\bibfnamefont {A.}~\bibnamefont
  {Linde}},\ }\href {\doibase 10.1088/1475-7516/2022/04/017} {\bibfield
  {journal} {\bibinfo  {journal} {JCAP}\ }\textbf {\bibinfo {volume} {04}},\
  \bibinfo {pages} {017} (\bibinfo {year} {2022})},\ \Eprint
  {http://arxiv.org/abs/2202.06492} {arXiv:2202.06492 [astro-ph.CO]}
  \BibitemShut {NoStop}%
\bibitem [{\citenamefont {Louis}\ \emph {et~al.}(2025)\citenamefont {Louis}
  \emph {et~al.}}]{ACT:2025fju}%
  \BibitemOpen
  \bibfield  {author} {\bibinfo {author} {\bibfnamefont {T.}~\bibnamefont
  {Louis}} \emph {et~al.} (\bibinfo {collaboration} {ACT}),\ }\href@noop {} {\
  (\bibinfo {year} {2025})},\ \Eprint {http://arxiv.org/abs/2503.14452}
  {arXiv:2503.14452 [astro-ph.CO]} \BibitemShut {NoStop}%
\bibitem [{\citenamefont {Calabrese}\ \emph {et~al.}(2025)\citenamefont
  {Calabrese} \emph {et~al.}}]{ACT:2025tim}%
  \BibitemOpen
  \bibfield  {author} {\bibinfo {author} {\bibfnamefont {E.}~\bibnamefont
  {Calabrese}} \emph {et~al.} (\bibinfo {collaboration} {ACT}),\ }\href@noop {}
  {\  (\bibinfo {year} {2025})},\ \Eprint {http://arxiv.org/abs/2503.14454}
  {arXiv:2503.14454 [astro-ph.CO]} \BibitemShut {NoStop}%
\bibitem [{\citenamefont {Kallosh}\ and\ \citenamefont
  {Linde}(2025)}]{Kallosh:2025ijd}%
  \BibitemOpen
  \bibfield  {author} {\bibinfo {author} {\bibfnamefont {R.}~\bibnamefont
  {Kallosh}}\ and\ \bibinfo {author} {\bibfnamefont {A.}~\bibnamefont
  {Linde}},\ }\href@noop {} {\  (\bibinfo {year} {2025})},\ \Eprint
  {http://arxiv.org/abs/2505.13646} {arXiv:2505.13646 [hep-th]} \BibitemShut
  {NoStop}%
\bibitem [{\citenamefont {Berera}(1995)}]{Berera:1995ie}%
  \BibitemOpen
  \bibfield  {author} {\bibinfo {author} {\bibfnamefont {A.}~\bibnamefont
  {Berera}},\ }\href {\doibase 10.1103/PhysRevLett.75.3218} {\bibfield
  {journal} {\bibinfo  {journal} {Phys. Rev. Lett.}\ }\textbf {\bibinfo
  {volume} {75}},\ \bibinfo {pages} {3218} (\bibinfo {year} {1995})},\ \Eprint
  {http://arxiv.org/abs/astro-ph/9509049} {arXiv:astro-ph/9509049} \BibitemShut
  {NoStop}%
\bibitem [{\citenamefont {Riotto}(2003)}]{Riotto:2002yw}%
  \BibitemOpen
  \bibfield  {author} {\bibinfo {author} {\bibfnamefont {A.}~\bibnamefont
  {Riotto}},\ }\href@noop {} {\bibfield  {journal} {\bibinfo  {journal} {ICTP
  Lect. Notes Ser.}\ }\textbf {\bibinfo {volume} {14}},\ \bibinfo {pages} {317}
  (\bibinfo {year} {2003})},\ \Eprint {http://arxiv.org/abs/hep-ph/0210162}
  {arXiv:hep-ph/0210162} \BibitemShut {NoStop}%
\bibitem [{\citenamefont {Baumann}(2011)}]{Baumann:2009ds}%
  \BibitemOpen
  \bibfield  {author} {\bibinfo {author} {\bibfnamefont {D.}~\bibnamefont
  {Baumann}},\ }in\ \href {\doibase 10.1142/9789814327183_0010} {\emph
  {\bibinfo {booktitle} {{Theoretical Advanced Study Institute in Elementary
  Particle Physics}: {Physics of the Large and the Small}}}}\ (\bibinfo {year}
  {2011})\ pp.\ \bibinfo {pages} {523--686},\ \Eprint
  {http://arxiv.org/abs/0907.5424} {arXiv:0907.5424 [hep-th]} \BibitemShut
  {NoStop}%
\bibitem [{\citenamefont {Kamali}\ \emph {et~al.}(2023)\citenamefont {Kamali},
  \citenamefont {Motaharfar},\ and\ \citenamefont {O.~Ramos}}]{Kamali:2023lzq}%
  \BibitemOpen
  \bibfield  {author} {\bibinfo {author} {\bibfnamefont {V.}~\bibnamefont
  {Kamali}}, \bibinfo {author} {\bibfnamefont {M.}~\bibnamefont {Motaharfar}},
  \ and\ \bibinfo {author} {\bibfnamefont {R.}~\bibnamefont {O.~Ramos}},\
  }\href {\doibase 10.3390/universe9030124} {\bibfield  {journal} {\bibinfo
  {journal} {Universe}\ }\textbf {\bibinfo {volume} {9}},\ \bibinfo {pages}
  {124} (\bibinfo {year} {2023})},\ \Eprint {http://arxiv.org/abs/2302.02827}
  {arXiv:2302.02827 [hep-ph]} \BibitemShut {NoStop}%
\bibitem [{\citenamefont {Berera}(2023)}]{Berera:2023liv}%
  \BibitemOpen
  \bibfield  {author} {\bibinfo {author} {\bibfnamefont {A.}~\bibnamefont
  {Berera}},\ }\href {\doibase 10.3390/universe9060272} {\bibfield  {journal}
  {\bibinfo  {journal} {Universe}\ }\textbf {\bibinfo {volume} {9}},\ \bibinfo
  {pages} {272} (\bibinfo {year} {2023})},\ \Eprint
  {http://arxiv.org/abs/2305.10879} {arXiv:2305.10879 [hep-ph]} \BibitemShut
  {NoStop}%
\bibitem [{\citenamefont {Bartrum}\ \emph {et~al.}(2014)\citenamefont
  {Bartrum}, \citenamefont {Bastero-Gil}, \citenamefont {Berera}, \citenamefont
  {Cerezo}, \citenamefont {Ramos},\ and\ \citenamefont
  {Rosa}}]{Bartrum:2013fia}%
  \BibitemOpen
  \bibfield  {author} {\bibinfo {author} {\bibfnamefont {S.}~\bibnamefont
  {Bartrum}}, \bibinfo {author} {\bibfnamefont {M.}~\bibnamefont
  {Bastero-Gil}}, \bibinfo {author} {\bibfnamefont {A.}~\bibnamefont {Berera}},
  \bibinfo {author} {\bibfnamefont {R.}~\bibnamefont {Cerezo}}, \bibinfo
  {author} {\bibfnamefont {R.~O.}\ \bibnamefont {Ramos}}, \ and\ \bibinfo
  {author} {\bibfnamefont {J.~G.}\ \bibnamefont {Rosa}},\ }\href {\doibase
  10.1016/j.physletb.2014.03.029} {\bibfield  {journal} {\bibinfo  {journal}
  {Phys. Lett. B}\ }\textbf {\bibinfo {volume} {732}},\ \bibinfo {pages} {116}
  (\bibinfo {year} {2014})},\ \Eprint {http://arxiv.org/abs/1307.5868}
  {arXiv:1307.5868 [hep-ph]} \BibitemShut {NoStop}%
\bibitem [{\citenamefont {Berera}\ \emph {et~al.}(2025)\citenamefont {Berera},
  \citenamefont {Brahma}, \citenamefont {Qiu}, \citenamefont {O.~Ramos},\ and\
  \citenamefont {Rodrigues}}]{Berera:2025vsu}%
  \BibitemOpen
  \bibfield  {author} {\bibinfo {author} {\bibfnamefont {A.}~\bibnamefont
  {Berera}}, \bibinfo {author} {\bibfnamefont {S.}~\bibnamefont {Brahma}},
  \bibinfo {author} {\bibfnamefont {Z.}~\bibnamefont {Qiu}}, \bibinfo {author}
  {\bibfnamefont {R.}~\bibnamefont {O.~Ramos}}, \ and\ \bibinfo {author}
  {\bibfnamefont {G.~S.}\ \bibnamefont {Rodrigues}},\ }\href@noop {} {\
  (\bibinfo {year} {2025})},\ \Eprint {http://arxiv.org/abs/2504.02655}
  {arXiv:2504.02655 [hep-th]} \BibitemShut {NoStop}%
\bibitem [{\citenamefont {Das}\ and\ \citenamefont
  {Ramos}(2020)}]{Das:2020xmh}%
  \BibitemOpen
  \bibfield  {author} {\bibinfo {author} {\bibfnamefont {S.}~\bibnamefont
  {Das}}\ and\ \bibinfo {author} {\bibfnamefont {R.~O.}\ \bibnamefont
  {Ramos}},\ }\href {\doibase 10.1103/PhysRevD.102.103522} {\bibfield
  {journal} {\bibinfo  {journal} {Phys. Rev. D}\ }\textbf {\bibinfo {volume}
  {102}},\ \bibinfo {pages} {103522} (\bibinfo {year} {2020})},\ \Eprint
  {http://arxiv.org/abs/2007.15268} {arXiv:2007.15268 [hep-th]} \BibitemShut
  {NoStop}%
\bibitem [{\citenamefont {Das}(2019{\natexlab{a}})}]{Das:2018hqy}%
  \BibitemOpen
  \bibfield  {author} {\bibinfo {author} {\bibfnamefont {S.}~\bibnamefont
  {Das}},\ }\href {\doibase 10.1103/PhysRevD.99.083510} {\bibfield  {journal}
  {\bibinfo  {journal} {Phys. Rev. D}\ }\textbf {\bibinfo {volume} {99}},\
  \bibinfo {pages} {083510} (\bibinfo {year} {2019}{\natexlab{a}})},\ \Eprint
  {http://arxiv.org/abs/1809.03962} {arXiv:1809.03962 [hep-th]} \BibitemShut
  {NoStop}%
\bibitem [{\citenamefont {Das}(2019{\natexlab{b}})}]{Das:2018rpg}%
  \BibitemOpen
  \bibfield  {author} {\bibinfo {author} {\bibfnamefont {S.}~\bibnamefont
  {Das}},\ }\href {\doibase 10.1103/PhysRevD.99.063514} {\bibfield  {journal}
  {\bibinfo  {journal} {Phys. Rev. D}\ }\textbf {\bibinfo {volume} {99}},\
  \bibinfo {pages} {063514} (\bibinfo {year} {2019}{\natexlab{b}})},\ \Eprint
  {http://arxiv.org/abs/1810.05038} {arXiv:1810.05038 [hep-th]} \BibitemShut
  {NoStop}%
\bibitem [{\citenamefont {Motaharfar}\ \emph {et~al.}(2019)\citenamefont
  {Motaharfar}, \citenamefont {Kamali},\ and\ \citenamefont
  {Ramos}}]{Motaharfar:2018zyb}%
  \BibitemOpen
  \bibfield  {author} {\bibinfo {author} {\bibfnamefont {M.}~\bibnamefont
  {Motaharfar}}, \bibinfo {author} {\bibfnamefont {V.}~\bibnamefont {Kamali}},
  \ and\ \bibinfo {author} {\bibfnamefont {R.~O.}\ \bibnamefont {Ramos}},\
  }\href {\doibase 10.1103/PhysRevD.99.063513} {\bibfield  {journal} {\bibinfo
  {journal} {Phys. Rev. D}\ }\textbf {\bibinfo {volume} {99}},\ \bibinfo
  {pages} {063513} (\bibinfo {year} {2019})},\ \Eprint
  {http://arxiv.org/abs/1810.02816} {arXiv:1810.02816 [astro-ph.CO]}
  \BibitemShut {NoStop}%
\bibitem [{\citenamefont {Das}(2020)}]{Das:2019hto}%
  \BibitemOpen
  \bibfield  {author} {\bibinfo {author} {\bibfnamefont {S.}~\bibnamefont
  {Das}},\ }\href {\doibase 10.1016/j.dark.2019.100432} {\bibfield  {journal}
  {\bibinfo  {journal} {Phys. Dark Univ.}\ }\textbf {\bibinfo {volume} {27}},\
  \bibinfo {pages} {100432} (\bibinfo {year} {2020})},\ \Eprint
  {http://arxiv.org/abs/1910.02147} {arXiv:1910.02147 [hep-th]} \BibitemShut
  {NoStop}%
\bibitem [{\citenamefont {Das}\ \emph {et~al.}(2020)\citenamefont {Das},
  \citenamefont {Goswami},\ and\ \citenamefont {Krishnan}}]{Das:2019acf}%
  \BibitemOpen
  \bibfield  {author} {\bibinfo {author} {\bibfnamefont {S.}~\bibnamefont
  {Das}}, \bibinfo {author} {\bibfnamefont {G.}~\bibnamefont {Goswami}}, \ and\
  \bibinfo {author} {\bibfnamefont {C.}~\bibnamefont {Krishnan}},\ }\href
  {\doibase 10.1103/PhysRevD.101.103529} {\bibfield  {journal} {\bibinfo
  {journal} {Phys. Rev. D}\ }\textbf {\bibinfo {volume} {101}},\ \bibinfo
  {pages} {103529} (\bibinfo {year} {2020})},\ \Eprint
  {http://arxiv.org/abs/1911.00323} {arXiv:1911.00323 [hep-th]} \BibitemShut
  {NoStop}%
\bibitem [{\citenamefont {Chakraborty}\ and\ \citenamefont
  {O.~Ramos}(2025)}]{Chakraborty:2025yms}%
  \BibitemOpen
  \bibfield  {author} {\bibinfo {author} {\bibfnamefont {D.}~\bibnamefont
  {Chakraborty}}\ and\ \bibinfo {author} {\bibfnamefont {R.}~\bibnamefont
  {O.~Ramos}},\ }\href@noop {} {\  (\bibinfo {year} {2025})},\ \Eprint
  {http://arxiv.org/abs/2505.04447} {arXiv:2505.04447 [hep-th]} \BibitemShut
  {NoStop}%
\bibitem [{\citenamefont {Bastero-Gil}\ \emph {et~al.}(2014)\citenamefont
  {Bastero-Gil}, \citenamefont {Berera}, \citenamefont {Moss},\ and\
  \citenamefont {Ramos}}]{Bastero-Gil:2014raa}%
  \BibitemOpen
  \bibfield  {author} {\bibinfo {author} {\bibfnamefont {M.}~\bibnamefont
  {Bastero-Gil}}, \bibinfo {author} {\bibfnamefont {A.}~\bibnamefont {Berera}},
  \bibinfo {author} {\bibfnamefont {I.~G.}\ \bibnamefont {Moss}}, \ and\
  \bibinfo {author} {\bibfnamefont {R.~O.}\ \bibnamefont {Ramos}},\ }\href
  {\doibase 10.1088/1475-7516/2014/12/008} {\bibfield  {journal} {\bibinfo
  {journal} {JCAP}\ }\textbf {\bibinfo {volume} {12}},\ \bibinfo {pages} {008}
  (\bibinfo {year} {2014})},\ \Eprint {http://arxiv.org/abs/1408.4391}
  {arXiv:1408.4391 [astro-ph.CO]} \BibitemShut {NoStop}%
\bibitem [{\citenamefont {Mirbabayi}\ and\ \citenamefont
  {Gruzinov}(2023)}]{Mirbabayi:2022cbt}%
  \BibitemOpen
  \bibfield  {author} {\bibinfo {author} {\bibfnamefont {M.}~\bibnamefont
  {Mirbabayi}}\ and\ \bibinfo {author} {\bibfnamefont {A.}~\bibnamefont
  {Gruzinov}},\ }\href {\doibase 10.1088/1475-7516/2023/02/012} {\bibfield
  {journal} {\bibinfo  {journal} {JCAP}\ }\textbf {\bibinfo {volume} {02}},\
  \bibinfo {pages} {012} (\bibinfo {year} {2023})},\ \Eprint
  {http://arxiv.org/abs/2205.13227} {arXiv:2205.13227 [astro-ph.CO]}
  \BibitemShut {NoStop}%
\bibitem [{\citenamefont {Hall}\ \emph {et~al.}(2004)\citenamefont {Hall},
  \citenamefont {Moss},\ and\ \citenamefont {Berera}}]{Hall:2003zp}%
  \BibitemOpen
  \bibfield  {author} {\bibinfo {author} {\bibfnamefont {L.~M.~H.}\
  \bibnamefont {Hall}}, \bibinfo {author} {\bibfnamefont {I.~G.}\ \bibnamefont
  {Moss}}, \ and\ \bibinfo {author} {\bibfnamefont {A.}~\bibnamefont
  {Berera}},\ }\href {\doibase 10.1103/PhysRevD.69.083525} {\bibfield
  {journal} {\bibinfo  {journal} {Phys. Rev. D}\ }\textbf {\bibinfo {volume}
  {69}},\ \bibinfo {pages} {083525} (\bibinfo {year} {2004})},\ \Eprint
  {http://arxiv.org/abs/astro-ph/0305015} {arXiv:astro-ph/0305015} \BibitemShut
  {NoStop}%
\bibitem [{\citenamefont {Graham}\ and\ \citenamefont
  {Moss}(2009)}]{Graham:2009bf}%
  \BibitemOpen
  \bibfield  {author} {\bibinfo {author} {\bibfnamefont {C.}~\bibnamefont
  {Graham}}\ and\ \bibinfo {author} {\bibfnamefont {I.~G.}\ \bibnamefont
  {Moss}},\ }\href {\doibase 10.1088/1475-7516/2009/07/013} {\bibfield
  {journal} {\bibinfo  {journal} {JCAP}\ }\textbf {\bibinfo {volume} {07}},\
  \bibinfo {pages} {013} (\bibinfo {year} {2009})},\ \Eprint
  {http://arxiv.org/abs/0905.3500} {arXiv:0905.3500 [astro-ph.CO]} \BibitemShut
  {NoStop}%
\bibitem [{\citenamefont {Bastero-Gil}\ \emph {et~al.}(2011)\citenamefont
  {Bastero-Gil}, \citenamefont {Berera},\ and\ \citenamefont
  {Ramos}}]{Bastero-Gil:2011rva}%
  \BibitemOpen
  \bibfield  {author} {\bibinfo {author} {\bibfnamefont {M.}~\bibnamefont
  {Bastero-Gil}}, \bibinfo {author} {\bibfnamefont {A.}~\bibnamefont {Berera}},
  \ and\ \bibinfo {author} {\bibfnamefont {R.~O.}\ \bibnamefont {Ramos}},\
  }\href {\doibase 10.1088/1475-7516/2011/07/030} {\bibfield  {journal}
  {\bibinfo  {journal} {JCAP}\ }\textbf {\bibinfo {volume} {07}},\ \bibinfo
  {pages} {030} (\bibinfo {year} {2011})},\ \Eprint
  {http://arxiv.org/abs/1106.0701} {arXiv:1106.0701 [astro-ph.CO]} \BibitemShut
  {NoStop}%
\bibitem [{\citenamefont {Ramos}\ and\ \citenamefont
  {da~Silva}(2013)}]{Ramos:2013nsa}%
  \BibitemOpen
  \bibfield  {author} {\bibinfo {author} {\bibfnamefont {R.~O.}\ \bibnamefont
  {Ramos}}\ and\ \bibinfo {author} {\bibfnamefont {L.~A.}\ \bibnamefont
  {da~Silva}},\ }\href {\doibase 10.1088/1475-7516/2013/03/032} {\bibfield
  {journal} {\bibinfo  {journal} {JCAP}\ }\textbf {\bibinfo {volume} {03}},\
  \bibinfo {pages} {032} (\bibinfo {year} {2013})},\ \Eprint
  {http://arxiv.org/abs/1302.3544} {arXiv:1302.3544 [astro-ph.CO]} \BibitemShut
  {NoStop}%
\bibitem [{\citenamefont {Das}\ and\ \citenamefont
  {O.~Ramos}(2023)}]{Das:2022ubr}%
  \BibitemOpen
  \bibfield  {author} {\bibinfo {author} {\bibfnamefont {S.}~\bibnamefont
  {Das}}\ and\ \bibinfo {author} {\bibfnamefont {R.}~\bibnamefont {O.~Ramos}},\
  }\href {\doibase 10.3390/universe9020076} {\bibfield  {journal} {\bibinfo
  {journal} {Universe}\ }\textbf {\bibinfo {volume} {9}},\ \bibinfo {pages}
  {76} (\bibinfo {year} {2023})},\ \Eprint {http://arxiv.org/abs/2212.13914}
  {arXiv:2212.13914 [astro-ph.CO]} \BibitemShut {NoStop}%
\bibitem [{\citenamefont {Berghaus}\ \emph {et~al.}(2020)\citenamefont
  {Berghaus}, \citenamefont {Graham},\ and\ \citenamefont
  {Kaplan}}]{Berghaus:2019whh}%
  \BibitemOpen
  \bibfield  {author} {\bibinfo {author} {\bibfnamefont {K.~V.}\ \bibnamefont
  {Berghaus}}, \bibinfo {author} {\bibfnamefont {P.~W.}\ \bibnamefont
  {Graham}}, \ and\ \bibinfo {author} {\bibfnamefont {D.~E.}\ \bibnamefont
  {Kaplan}},\ }\href {\doibase 10.1088/1475-7516/2020/03/034} {\bibfield
  {journal} {\bibinfo  {journal} {JCAP}\ }\textbf {\bibinfo {volume} {03}},\
  \bibinfo {pages} {034} (\bibinfo {year} {2020})},\ \bibinfo {note} {[Erratum:
  JCAP 10, E02 (2023)]},\ \Eprint {http://arxiv.org/abs/1910.07525}
  {arXiv:1910.07525 [hep-ph]} \BibitemShut {NoStop}%
\bibitem [{\citenamefont {Das}\ and\ \citenamefont
  {O.~Ramos}(2021)}]{Das:2020lut}%
  \BibitemOpen
  \bibfield  {author} {\bibinfo {author} {\bibfnamefont {S.}~\bibnamefont
  {Das}}\ and\ \bibinfo {author} {\bibfnamefont {R.}~\bibnamefont {O.~Ramos}},\
  }\href {\doibase 10.1103/PhysRevD.103.123520} {\bibfield  {journal} {\bibinfo
   {journal} {Phys. Rev. D}\ }\textbf {\bibinfo {volume} {103}},\ \bibinfo
  {pages} {123520} (\bibinfo {year} {2021})},\ \Eprint
  {http://arxiv.org/abs/2005.01122} {arXiv:2005.01122 [gr-qc]} \BibitemShut
  {NoStop}%
\bibitem [{\citenamefont {Starobinsky}(1980)}]{Starobinsky:1980te}%
  \BibitemOpen
  \bibfield  {author} {\bibinfo {author} {\bibfnamefont {A.~A.}\ \bibnamefont
  {Starobinsky}},\ }\href {\doibase 10.1016/0370-2693(80)90670-X} {\bibfield
  {journal} {\bibinfo  {journal} {Phys. Lett. B}\ }\textbf {\bibinfo {volume}
  {91}},\ \bibinfo {pages} {99} (\bibinfo {year} {1980})}\BibitemShut {NoStop}%
\bibitem [{\citenamefont {Bezrukov}\ and\ \citenamefont
  {Shaposhnikov}(2008)}]{Bezrukov:2007ep}%
  \BibitemOpen
  \bibfield  {author} {\bibinfo {author} {\bibfnamefont {F.~L.}\ \bibnamefont
  {Bezrukov}}\ and\ \bibinfo {author} {\bibfnamefont {M.}~\bibnamefont
  {Shaposhnikov}},\ }\href {\doibase 10.1016/j.physletb.2007.11.072} {\bibfield
   {journal} {\bibinfo  {journal} {Phys. Lett. B}\ }\textbf {\bibinfo {volume}
  {659}},\ \bibinfo {pages} {703} (\bibinfo {year} {2008})},\ \Eprint
  {http://arxiv.org/abs/0710.3755} {arXiv:0710.3755 [hep-th]} \BibitemShut
  {NoStop}%
\bibitem [{\citenamefont {Okada}\ \emph {et~al.}(2010)\citenamefont {Okada},
  \citenamefont {Rehman},\ and\ \citenamefont {Shafi}}]{Okada:2010jf}%
  \BibitemOpen
  \bibfield  {author} {\bibinfo {author} {\bibfnamefont {N.}~\bibnamefont
  {Okada}}, \bibinfo {author} {\bibfnamefont {M.~U.}\ \bibnamefont {Rehman}}, \
  and\ \bibinfo {author} {\bibfnamefont {Q.}~\bibnamefont {Shafi}},\ }\href
  {\doibase 10.1103/PhysRevD.82.043502} {\bibfield  {journal} {\bibinfo
  {journal} {Phys. Rev. D}\ }\textbf {\bibinfo {volume} {82}},\ \bibinfo
  {pages} {043502} (\bibinfo {year} {2010})},\ \Eprint
  {http://arxiv.org/abs/1005.5161} {arXiv:1005.5161 [hep-ph]} \BibitemShut
  {NoStop}%
\bibitem [{\citenamefont {Linde}\ \emph {et~al.}(2011)\citenamefont {Linde},
  \citenamefont {Noorbala},\ and\ \citenamefont {Westphal}}]{Linde:2011nh}%
  \BibitemOpen
  \bibfield  {author} {\bibinfo {author} {\bibfnamefont {A.}~\bibnamefont
  {Linde}}, \bibinfo {author} {\bibfnamefont {M.}~\bibnamefont {Noorbala}}, \
  and\ \bibinfo {author} {\bibfnamefont {A.}~\bibnamefont {Westphal}},\ }\href
  {\doibase 10.1088/1475-7516/2011/03/013} {\bibfield  {journal} {\bibinfo
  {journal} {JCAP}\ }\textbf {\bibinfo {volume} {03}},\ \bibinfo {pages} {013}
  (\bibinfo {year} {2011})},\ \Eprint {http://arxiv.org/abs/1101.2652}
  {arXiv:1101.2652 [hep-th]} \BibitemShut {NoStop}%
\bibitem [{\citenamefont {Ellis}\ \emph
  {et~al.}(2013{\natexlab{a}})\citenamefont {Ellis}, \citenamefont
  {Nanopoulos},\ and\ \citenamefont {Olive}}]{Ellis:2013xoa}%
  \BibitemOpen
  \bibfield  {author} {\bibinfo {author} {\bibfnamefont {J.}~\bibnamefont
  {Ellis}}, \bibinfo {author} {\bibfnamefont {D.~V.}\ \bibnamefont
  {Nanopoulos}}, \ and\ \bibinfo {author} {\bibfnamefont {K.~A.}\ \bibnamefont
  {Olive}},\ }\href {\doibase 10.1103/PhysRevLett.111.111301} {\bibfield
  {journal} {\bibinfo  {journal} {Phys. Rev. Lett.}\ }\textbf {\bibinfo
  {volume} {111}},\ \bibinfo {pages} {111301} (\bibinfo {year}
  {2013}{\natexlab{a}})},\ \bibinfo {note} {[Erratum: Phys.Rev.Lett. 111,
  129902 (2013)]},\ \Eprint {http://arxiv.org/abs/1305.1247} {arXiv:1305.1247
  [hep-th]} \BibitemShut {NoStop}%
\bibitem [{\citenamefont {Buchmuller}\ \emph {et~al.}(2013)\citenamefont
  {Buchmuller}, \citenamefont {Domcke},\ and\ \citenamefont
  {Kamada}}]{Buchmuller:2013zfa}%
  \BibitemOpen
  \bibfield  {author} {\bibinfo {author} {\bibfnamefont {W.}~\bibnamefont
  {Buchmuller}}, \bibinfo {author} {\bibfnamefont {V.}~\bibnamefont {Domcke}},
  \ and\ \bibinfo {author} {\bibfnamefont {K.}~\bibnamefont {Kamada}},\ }\href
  {\doibase 10.1016/j.physletb.2013.08.042} {\bibfield  {journal} {\bibinfo
  {journal} {Phys. Lett. B}\ }\textbf {\bibinfo {volume} {726}},\ \bibinfo
  {pages} {467} (\bibinfo {year} {2013})},\ \Eprint
  {http://arxiv.org/abs/1306.3471} {arXiv:1306.3471 [hep-th]} \BibitemShut
  {NoStop}%
\bibitem [{\citenamefont {Kallosh}\ and\ \citenamefont
  {Linde}(2013{\natexlab{b}})}]{Kallosh:2013daa}%
  \BibitemOpen
  \bibfield  {author} {\bibinfo {author} {\bibfnamefont {R.}~\bibnamefont
  {Kallosh}}\ and\ \bibinfo {author} {\bibfnamefont {A.}~\bibnamefont
  {Linde}},\ }\href {\doibase 10.1088/1475-7516/2013/12/006} {\bibfield
  {journal} {\bibinfo  {journal} {JCAP}\ }\textbf {\bibinfo {volume} {12}},\
  \bibinfo {pages} {006} (\bibinfo {year} {2013}{\natexlab{b}})},\ \Eprint
  {http://arxiv.org/abs/1309.2015} {arXiv:1309.2015 [hep-th]} \BibitemShut
  {NoStop}%
\bibitem [{\citenamefont {Kallosh}\ and\ \citenamefont
  {Linde}(2013{\natexlab{c}})}]{Kallosh:2013pby}%
  \BibitemOpen
  \bibfield  {author} {\bibinfo {author} {\bibfnamefont {R.}~\bibnamefont
  {Kallosh}}\ and\ \bibinfo {author} {\bibfnamefont {A.}~\bibnamefont
  {Linde}},\ }\href {\doibase 10.1088/1475-7516/2013/06/027} {\bibfield
  {journal} {\bibinfo  {journal} {JCAP}\ }\textbf {\bibinfo {volume} {06}},\
  \bibinfo {pages} {027} (\bibinfo {year} {2013}{\natexlab{c}})},\ \Eprint
  {http://arxiv.org/abs/1306.3211} {arXiv:1306.3211 [hep-th]} \BibitemShut
  {NoStop}%
\bibitem [{\citenamefont {Kallosh}\ and\ \citenamefont
  {Linde}(2013{\natexlab{d}})}]{Kallosh:2013lkr}%
  \BibitemOpen
  \bibfield  {author} {\bibinfo {author} {\bibfnamefont {R.}~\bibnamefont
  {Kallosh}}\ and\ \bibinfo {author} {\bibfnamefont {A.}~\bibnamefont
  {Linde}},\ }\href {\doibase 10.1088/1475-7516/2013/06/028} {\bibfield
  {journal} {\bibinfo  {journal} {JCAP}\ }\textbf {\bibinfo {volume} {06}},\
  \bibinfo {pages} {028} (\bibinfo {year} {2013}{\natexlab{d}})},\ \Eprint
  {http://arxiv.org/abs/1306.3214} {arXiv:1306.3214 [hep-th]} \BibitemShut
  {NoStop}%
\bibitem [{\citenamefont {Ellis}\ \emph
  {et~al.}(2013{\natexlab{b}})\citenamefont {Ellis}, \citenamefont
  {Nanopoulos},\ and\ \citenamefont {Olive}}]{Ellis:2013nxa}%
  \BibitemOpen
  \bibfield  {author} {\bibinfo {author} {\bibfnamefont {J.}~\bibnamefont
  {Ellis}}, \bibinfo {author} {\bibfnamefont {D.~V.}\ \bibnamefont
  {Nanopoulos}}, \ and\ \bibinfo {author} {\bibfnamefont {K.~A.}\ \bibnamefont
  {Olive}},\ }\href {\doibase 10.1088/1475-7516/2013/10/009} {\bibfield
  {journal} {\bibinfo  {journal} {JCAP}\ }\textbf {\bibinfo {volume} {10}},\
  \bibinfo {pages} {009} (\bibinfo {year} {2013}{\natexlab{b}})},\ \Eprint
  {http://arxiv.org/abs/1307.3537} {arXiv:1307.3537 [hep-th]} \BibitemShut
  {NoStop}%
\bibitem [{\citenamefont {Ferrara}\ \emph {et~al.}(2013)\citenamefont
  {Ferrara}, \citenamefont {Kallosh}, \citenamefont {Linde},\ and\
  \citenamefont {Porrati}}]{Ferrara:2013rsa}%
  \BibitemOpen
  \bibfield  {author} {\bibinfo {author} {\bibfnamefont {S.}~\bibnamefont
  {Ferrara}}, \bibinfo {author} {\bibfnamefont {R.}~\bibnamefont {Kallosh}},
  \bibinfo {author} {\bibfnamefont {A.}~\bibnamefont {Linde}}, \ and\ \bibinfo
  {author} {\bibfnamefont {M.}~\bibnamefont {Porrati}},\ }\href {\doibase
  10.1103/PhysRevD.88.085038} {\bibfield  {journal} {\bibinfo  {journal} {Phys.
  Rev. D}\ }\textbf {\bibinfo {volume} {88}},\ \bibinfo {pages} {085038}
  (\bibinfo {year} {2013})},\ \Eprint {http://arxiv.org/abs/1307.7696}
  {arXiv:1307.7696 [hep-th]} \BibitemShut {NoStop}%
\bibitem [{\citenamefont {Kallosh}\ and\ \citenamefont
  {Linde}(2013{\natexlab{e}})}]{Kallosh:2013maa}%
  \BibitemOpen
  \bibfield  {author} {\bibinfo {author} {\bibfnamefont {R.}~\bibnamefont
  {Kallosh}}\ and\ \bibinfo {author} {\bibfnamefont {A.}~\bibnamefont
  {Linde}},\ }\href {\doibase 10.1088/1475-7516/2013/10/033} {\bibfield
  {journal} {\bibinfo  {journal} {JCAP}\ }\textbf {\bibinfo {volume} {10}},\
  \bibinfo {pages} {033} (\bibinfo {year} {2013}{\natexlab{e}})},\ \Eprint
  {http://arxiv.org/abs/1307.7938} {arXiv:1307.7938 [hep-th]} \BibitemShut
  {NoStop}%
\bibitem [{\citenamefont {Kallosh}\ \emph {et~al.}(2014)\citenamefont
  {Kallosh}, \citenamefont {Linde},\ and\ \citenamefont
  {Roest}}]{Kallosh:2013tua}%
  \BibitemOpen
  \bibfield  {author} {\bibinfo {author} {\bibfnamefont {R.}~\bibnamefont
  {Kallosh}}, \bibinfo {author} {\bibfnamefont {A.}~\bibnamefont {Linde}}, \
  and\ \bibinfo {author} {\bibfnamefont {D.}~\bibnamefont {Roest}},\ }\href
  {\doibase 10.1103/PhysRevLett.112.011303} {\bibfield  {journal} {\bibinfo
  {journal} {Phys. Rev. Lett.}\ }\textbf {\bibinfo {volume} {112}},\ \bibinfo
  {pages} {011303} (\bibinfo {year} {2014})},\ \Eprint
  {http://arxiv.org/abs/1310.3950} {arXiv:1310.3950 [hep-th]} \BibitemShut
  {NoStop}%
\bibitem [{\citenamefont {Roest}(2014)}]{Roest:2013fha}%
  \BibitemOpen
  \bibfield  {author} {\bibinfo {author} {\bibfnamefont {D.}~\bibnamefont
  {Roest}},\ }\href {\doibase 10.1088/1475-7516/2014/01/007} {\bibfield
  {journal} {\bibinfo  {journal} {JCAP}\ }\textbf {\bibinfo {volume} {01}},\
  \bibinfo {pages} {007} (\bibinfo {year} {2014})},\ \Eprint
  {http://arxiv.org/abs/1309.1285} {arXiv:1309.1285 [hep-th]} \BibitemShut
  {NoStop}%
\bibitem [{\citenamefont {Akrami}\ \emph {et~al.}(2020)\citenamefont {Akrami}
  \emph {et~al.}}]{Planck:2018jri}%
  \BibitemOpen
  \bibfield  {author} {\bibinfo {author} {\bibfnamefont {Y.}~\bibnamefont
  {Akrami}} \emph {et~al.} (\bibinfo {collaboration} {Planck}),\ }\href
  {\doibase 10.1051/0004-6361/201833887} {\bibfield  {journal} {\bibinfo
  {journal} {Astron. Astrophys.}\ }\textbf {\bibinfo {volume} {641}},\ \bibinfo
  {pages} {A10} (\bibinfo {year} {2020})},\ \Eprint
  {http://arxiv.org/abs/1807.06211} {arXiv:1807.06211 [astro-ph.CO]}
  \BibitemShut {NoStop}%
\bibitem [{\citenamefont {Ballardini}(2024)}]{Ballardini:2024ado}%
  \BibitemOpen
  \bibfield  {author} {\bibinfo {author} {\bibfnamefont {M.}~\bibnamefont
  {Ballardini}},\ }\href@noop {} {\  (\bibinfo {year} {2024})},\ \Eprint
  {http://arxiv.org/abs/2408.03321} {arXiv:2408.03321 [astro-ph.CO]}
  \BibitemShut {NoStop}%
\bibitem [{\citenamefont {Dimopoulos}\ and\ \citenamefont
  {Owen}(2016)}]{Dimopoulos:2016zhy}%
  \BibitemOpen
  \bibfield  {author} {\bibinfo {author} {\bibfnamefont {K.}~\bibnamefont
  {Dimopoulos}}\ and\ \bibinfo {author} {\bibfnamefont {C.}~\bibnamefont
  {Owen}},\ }\href {\doibase 10.1103/PhysRevD.94.063518} {\bibfield  {journal}
  {\bibinfo  {journal} {Phys. Rev. D}\ }\textbf {\bibinfo {volume} {94}},\
  \bibinfo {pages} {063518} (\bibinfo {year} {2016})},\ \Eprint
  {http://arxiv.org/abs/1607.02469} {arXiv:1607.02469 [hep-ph]} \BibitemShut
  {NoStop}%
\bibitem [{\citenamefont {Adhikari}\ \emph {et~al.}(2020)\citenamefont
  {Adhikari}, \citenamefont {Gangopadhyay},\ and\ \citenamefont
  {Yogesh}}]{Adhikari:2020xcg}%
  \BibitemOpen
  \bibfield  {author} {\bibinfo {author} {\bibfnamefont {R.}~\bibnamefont
  {Adhikari}}, \bibinfo {author} {\bibfnamefont {M.~R.}\ \bibnamefont
  {Gangopadhyay}}, \ and\ \bibinfo {author} {\bibnamefont {Yogesh}},\ }\href
  {\doibase 10.1140/epjc/s10052-020-08460-3} {\bibfield  {journal} {\bibinfo
  {journal} {Eur. Phys. J. C}\ }\textbf {\bibinfo {volume} {80}},\ \bibinfo
  {pages} {899} (\bibinfo {year} {2020})},\ \Eprint
  {http://arxiv.org/abs/2002.07061} {arXiv:2002.07061 [astro-ph.CO]}
  \BibitemShut {NoStop}%
\bibitem [{\citenamefont {Bhattacharya}\ \emph {et~al.}(2023)\citenamefont
  {Bhattacharya}, \citenamefont {Dutta}, \citenamefont {Gangopadhyay},\ and\
  \citenamefont {Maharana}}]{Bhattacharya:2022akq}%
  \BibitemOpen
  \bibfield  {author} {\bibinfo {author} {\bibfnamefont {S.}~\bibnamefont
  {Bhattacharya}}, \bibinfo {author} {\bibfnamefont {K.}~\bibnamefont {Dutta}},
  \bibinfo {author} {\bibfnamefont {M.~R.}\ \bibnamefont {Gangopadhyay}}, \
  and\ \bibinfo {author} {\bibfnamefont {A.}~\bibnamefont {Maharana}},\ }\href
  {\doibase 10.1103/PhysRevD.107.103530} {\bibfield  {journal} {\bibinfo
  {journal} {Phys. Rev. D}\ }\textbf {\bibinfo {volume} {107}},\ \bibinfo
  {pages} {103530} (\bibinfo {year} {2023})},\ \Eprint
  {http://arxiv.org/abs/2212.13363} {arXiv:2212.13363 [astro-ph.CO]}
  \BibitemShut {NoStop}%
\bibitem [{\citenamefont {Ballesteros}\ \emph {et~al.}(2024)\citenamefont
  {Ballesteros}, \citenamefont {Perez~Rodriguez},\ and\ \citenamefont
  {Pierre}}]{Ballesteros:2023dno}%
  \BibitemOpen
  \bibfield  {author} {\bibinfo {author} {\bibfnamefont {G.}~\bibnamefont
  {Ballesteros}}, \bibinfo {author} {\bibfnamefont {A.}~\bibnamefont
  {Perez~Rodriguez}}, \ and\ \bibinfo {author} {\bibfnamefont {M.}~\bibnamefont
  {Pierre}},\ }\href {\doibase 10.1088/1475-7516/2024/03/003} {\bibfield
  {journal} {\bibinfo  {journal} {JCAP}\ }\textbf {\bibinfo {volume} {03}},\
  \bibinfo {pages} {003} (\bibinfo {year} {2024})},\ \Eprint
  {http://arxiv.org/abs/2304.05978} {arXiv:2304.05978 [astro-ph.CO]}
  \BibitemShut {NoStop}%
\bibitem [{\citenamefont {Kumar}\ and\ \citenamefont
  {Das}(2024)}]{Kumar:2024hju}%
  \BibitemOpen
  \bibfield  {author} {\bibinfo {author} {\bibfnamefont {U.}~\bibnamefont
  {Kumar}}\ and\ \bibinfo {author} {\bibfnamefont {S.}~\bibnamefont {Das}},\
  }\href {\doibase 10.1088/1475-7516/2024/10/058} {\bibfield  {journal}
  {\bibinfo  {journal} {JCAP}\ }\textbf {\bibinfo {volume} {10}},\ \bibinfo
  {pages} {058} (\bibinfo {year} {2024})},\ \Eprint
  {http://arxiv.org/abs/2407.06032} {arXiv:2407.06032 [astro-ph.CO]}
  \BibitemShut {NoStop}%
\bibitem [{\citenamefont {Yamada}(2018)}]{Yamada:2018nsk}%
  \BibitemOpen
  \bibfield  {author} {\bibinfo {author} {\bibfnamefont {Y.}~\bibnamefont
  {Yamada}},\ }\href {\doibase 10.1007/JHEP04(2018)006} {\bibfield  {journal}
  {\bibinfo  {journal} {JHEP}\ }\textbf {\bibinfo {volume} {04}},\ \bibinfo
  {pages} {006} (\bibinfo {year} {2018})},\ \Eprint
  {http://arxiv.org/abs/1802.04848} {arXiv:1802.04848 [hep-th]} \BibitemShut
  {NoStop}%
\bibitem [{\citenamefont {Rodrigues}\ and\ \citenamefont
  {O.~Ramos}(2025)}]{Rodrigues:2025neh}%
  \BibitemOpen
  \bibfield  {author} {\bibinfo {author} {\bibfnamefont {G.~S.}\ \bibnamefont
  {Rodrigues}}\ and\ \bibinfo {author} {\bibfnamefont {R.}~\bibnamefont
  {O.~Ramos}},\ }\href@noop {} {\  (\bibinfo {year} {2025})},\ \Eprint
  {http://arxiv.org/abs/2504.17760} {arXiv:2504.17760 [astro-ph.CO]}
  \BibitemShut {NoStop}%
\bibitem [{\citenamefont {Montefalcone}\ \emph {et~al.}(2024)\citenamefont
  {Montefalcone}, \citenamefont {Aragam}, \citenamefont {Visinelli},\ and\
  \citenamefont {Freese}}]{Montefalcone:2023pvh}%
  \BibitemOpen
  \bibfield  {author} {\bibinfo {author} {\bibfnamefont {G.}~\bibnamefont
  {Montefalcone}}, \bibinfo {author} {\bibfnamefont {V.}~\bibnamefont
  {Aragam}}, \bibinfo {author} {\bibfnamefont {L.}~\bibnamefont {Visinelli}}, \
  and\ \bibinfo {author} {\bibfnamefont {K.}~\bibnamefont {Freese}},\ }\href
  {\doibase 10.1088/1475-7516/2024/01/032} {\bibfield  {journal} {\bibinfo
  {journal} {JCAP}\ }\textbf {\bibinfo {volume} {01}},\ \bibinfo {pages} {032}
  (\bibinfo {year} {2024})},\ \Eprint {http://arxiv.org/abs/2306.16190}
  {arXiv:2306.16190 [astro-ph.CO]} \BibitemShut {NoStop}%
\bibitem [{\citenamefont {Allys}\ \emph {et~al.}(2023)\citenamefont {Allys}
  \emph {et~al.}}]{LiteBIRD:2022cnt}%
  \BibitemOpen
  \bibfield  {author} {\bibinfo {author} {\bibfnamefont {E.}~\bibnamefont
  {Allys}} \emph {et~al.} (\bibinfo {collaboration} {LiteBIRD}),\ }\href
  {\doibase 10.1093/ptep/ptac150} {\bibfield  {journal} {\bibinfo  {journal}
  {PTEP}\ }\textbf {\bibinfo {volume} {2023}},\ \bibinfo {pages} {042F01}
  (\bibinfo {year} {2023})},\ \Eprint {http://arxiv.org/abs/2202.02773}
  {arXiv:2202.02773 [astro-ph.IM]} \BibitemShut {NoStop}%
\bibitem [{\citenamefont {Ade}\ \emph {et~al.}(2019)\citenamefont {Ade} \emph
  {et~al.}}]{SimonsObservatory:2018koc}%
  \BibitemOpen
  \bibfield  {author} {\bibinfo {author} {\bibfnamefont {P.}~\bibnamefont
  {Ade}} \emph {et~al.} (\bibinfo {collaboration} {Simons Observatory}),\
  }\href {\doibase 10.1088/1475-7516/2019/02/056} {\bibfield  {journal}
  {\bibinfo  {journal} {JCAP}\ }\textbf {\bibinfo {volume} {02}},\ \bibinfo
  {pages} {056} (\bibinfo {year} {2019})},\ \Eprint
  {http://arxiv.org/abs/1808.07445} {arXiv:1808.07445 [astro-ph.CO]}
  \BibitemShut {NoStop}%
\bibitem [{\citenamefont {Saha}\ and\ \citenamefont
  {Nandy}(2025)}]{Saha:2025yzf}%
  \BibitemOpen
  \bibfield  {author} {\bibinfo {author} {\bibfnamefont {B.}~\bibnamefont
  {Saha}}\ and\ \bibinfo {author} {\bibfnamefont {M.~K.}\ \bibnamefont
  {Nandy}},\ }\href@noop {} {\  (\bibinfo {year} {2025})},\ \Eprint
  {http://arxiv.org/abs/2506.06717} {arXiv:2506.06717 [astro-ph.CO]}
  \BibitemShut {NoStop}%
\end{thebibliography}%


\end{document}